\documentclass[a4paper, 10pt, oneside, onecolumn, preprint]{elsarticle}

\usepackage{amsmath,amssymb}
\usepackage{graphicx}
\usepackage{subfigure}
\usepackage{bm}
\usepackage{color}
\usepackage[colorlinks,citecolor=blue]{hyperref}
\usepackage{amsfonts}
\usepackage{dcolumn}
\usepackage{times}

\begin{document}
\title{Symmetry protected topological phases in spin-$1$ ladders and their phase transitions}
\author[Chen]{Ji-Yao Chen\corref{cor1}}
\ead{chenjiyao11@mails.tsinghua.edu.cn}

\author[Liu]{Zheng-Xin Liu}
\ead{liuzxqh@mail.tsinghua.edu.cn}

\cortext[cor1]{Corresponding author}

\address[Chen]{State Key Laboratory of Low Dimensional Quantum Physics, Department of Physics, Tsinghua University, Beijing, 100084, P. R. China}
\address[Liu]{Institute for Advanced Study, Tsinghua University, Beijing, 100084, P. R. China}

\date{\today}
\begin{abstract}

We study two-legged spin-1 ladder systems with $D_2\times \sigma$ symmetry group, where $D_2$ is discrete spin rotational symmetry and $\sigma$ means interchain reflection symmetry. The system has one trivial phase and seven nontrivial symmetry protected topological (SPT) phases. We construct Hamiltonians to realize all of these SPT phases and study the phase transitions between them. Our numerical results indicate that there is no direct continuous transition between any two SPT phases we studied. We interpret our results via topological nonlinear sigma model effective field theory, and further conjecture that generally there is no direct continuous transition between two SPT phases in one dimension if the symmetry group is discrete at all length scales.

\end{abstract}


\maketitle

\section{Introduction}
Symmetry protected topological (SPT) phases are formed by gapped short-range-entangled quantum states that do not break any symmetry \cite{Zheng-Cheng Gu}. Contrary to trivial symmetric states, a nontrivial SPT state cannot be transformed into direct product state (or Slater determinant state for fermions) of local atomic basis via symmetric local unitary transformations. Nontrivial SPT states are characterized by their gapless or degenerate edge states at open boundaries, which are protected if the symmetry is reserved. On the other hand, when the symmetry is explicitly broken by perturbations, the nontrivial SPT phases can be adiabatically connected to the trivial phase and the boundary can be gapped out (for example, see \cite{pollmann 1}). The Haldane phase \cite{Haldane} in spin-1 chain is a one-dimensional (1D) bosonic SPT phase protected by $D_2$ spin rotation symmetry, or time-reversal symmetry, or spatial inversion symmetry \cite{pollmann 1, pollmann 2}. 1D bosonic SPT phases with onsite symmetry are classified by the projective representations of the symmetry group \cite{ChenGuWen11,ChenGuWen12,Schuch}. In higher dimensions, bosonic SPT phases are classified by group cohomology theory \cite{wen}, which is consistent with the continuous nonlinear sigma models \cite{CXu} or Chern-Simons theory (in two dimensions)\cite{yuanming lu}. According to the classification theory, new 1D SPT phases other than the Haldane phase have been constructed in spin chain/ladder models \cite{LiuLiuWen, LiuChenWen, spin one half}. It was also shown that different SPT phases can be distinguished by different responses of their edge states to external symmetry breaking fields. In 2D, the boundary of a nontrivial SPT phase is either gapless or symmetry breaking \cite{ChenLiuWen, LevinGu}. Nontrivial 2D SPT phases, such as integer bose/spin quantum Hall phases\cite{Senthil1, Senthil2, Furukawa, Jain, LuLee, YeWen, LiuMeiYeWen} and bosonic topological insulators \cite{XuSenthil,LiuGuWen}, have been realized. Important progress has also been made in classification and realization of SPT phases in three- or higher dimensions \cite{SenthilAshvin, Chen, WangSenthil, YeGu, JuvenWen}.

In the present work, we will study the model realization of 1D SPT phases in spin-1 ladder systems and the phase transitions between these phases. The symmetry group we are considering is an Abelian discrete group $D_2\times\sigma$. Here the $D_2$ subgroup is a discrete spin rotation symmetry $D_2 = \{E,R_x, R_y, R_z\}$ and $\sigma = \{E,P\}$ is the interchain exchange symmetry, where $E$ is the identity operation, $R_x(R_y, R_z)$ is a spin rotation of angle $\pi$ along $x(y,z)$ direction, and $P$ is interchain reflection. Since $\mathcal H^2(D_2\times\sigma,U(1))=\mathbb Z_2^3$, there are eight classes of projective representations, one is trivial and the others are nontrivial. Consequently there are seven distinct nontrivial SPT phases. The three root SPT phases according to the $\mathbb Z_2^3$ classification are called $t_x, t_y, t_z$, respectively, and all the other phases $t_{xy}=t_x\times t_y,\  t_{xz}=t_x\times t_z,\  t_{yz}=t_y\times t_z,\  t_0=t_x\times t_y\times t_z, \ I=t_x\times t_x=t_y\times t_y=t_z\times t_z$ can be generated by stacking two or more of the root phases (where the $t_0$ phase is the usual Haldane phase and $I$ is the trivial phase). Four of the nontrivial SPT phases, ${\it i.e.}$, $t_0, t_x, t_y, t_z$ have been constructed in spin-$\frac{1}{2}$ ladder systems \cite{spin one half}. However, due to a limited Hilbert space, the other three nontrivial phases cannot be realized with spin-$\frac{1}{2}$ ladders. Here we will realize all the eight SPT phases in spin-$1$ ladder models.

\begin{figure}
    \centering
    \includegraphics[width=8cm,height=6cm]{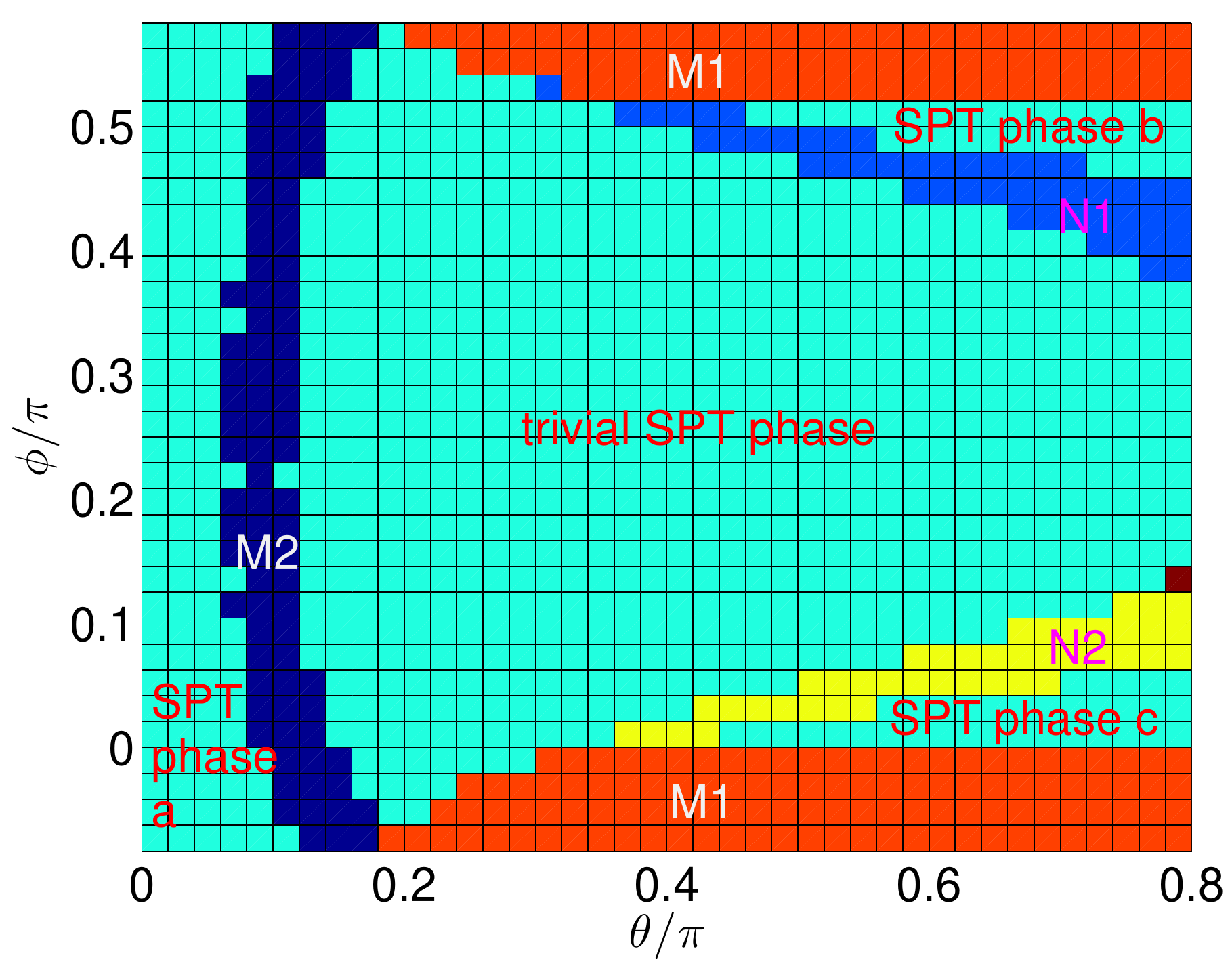}
\caption{Schematic phase diagram for the Hamiltonian (\ref{2d_PD_Ham}). There are four SPT phases, one of which is trivial. The SPT phases $a,b,c$ stand for $t_{xz}$, $t_y$ and $t_0$ phases respectively. The  symmetry breaking phases $\mathrm{M_1}$, $\mathrm{M_2}$, $\mathrm{N_1}$, $\mathrm{N_2}$  have order parameters $|\langle S^{y+}\rangle|$, $|\langle S^{y-}\rangle|$, $|\langle S^{z-}\rangle|$ and $|\langle S^{z+}\rangle|$ respectively. For details see Sec.~\ref{sec:trans}. }
\label{phasediagram}
\end{figure}

The phase transition between different SPT phases is an important issue. It is known that in spin-1 chain there is a second order phase transition between the Haldane phase and the trivial phase if the Hamiltonian has a continuous symmetry, ${\it e.g.}$, $U(1)$ spin rotation symmetry\cite{String}. On the other hand, if the symmetry group is discrete, only first order phase transitions between different SPT phases were observed in the literature \cite{Zheng-Cheng Gu, pollmann 1, LiuLiuWen}. In the following we study the transitions between different SPT phases and our results strongly support the following statement: in 1D a direct continuous transition between two SPT phases cannot take place if the symmetry group is discrete throughout. We will discuss about the special case where emergent continuous symmetry protects the continuous transition between two SPT phases (see \cite{DHLee2015}). 


A typical phase diagram of our model is shown in Fig.~\ref{phasediagram}, where $\theta, \phi$ are the parameters of the Hamiltonian [see Eq.(\ref{2d_PD_Ham})] and there are four SPT phases, one is trivial and the other three are nontrivial. Any two of these SPT phases are separated by at least one symmetry breaking phase. Direct phase transitions must be first order except that the symmetry group is enhanced to a continuous one. Our study will shed light on phase transitions between SPT phases with different symmetry groups and in higher dimensions.

The remaining part of the paper is organized as follows. In Sec.~\ref{sec:H}, we introduce the linear and projective representations of the symmetry group $D_2\times\sigma$, and give the active operators for each SPT phase. Further, we provide the Hamiltonian for each SPT phase. In Sec.~\ref{sec:trans}, we study the phase transitions between different SPT phases, and some of the results are explained in Sec.~\ref{discussPT} through nonlinear sigma model effective field theory. Sec.~\ref{sec:sum} is devoted to the conclusion.

\section{Parent Hamiltonian for each SPT phase}\label{sec:H}

Here we consider $S=1$ spin ladders respecting $D_2\times\sigma$ symmetry. The discrete Abelian group $D_2\times\sigma$ has eight one dimensional linear representations (see Table \ref{linear representation} of \ref{app:Rep}). The two $S=1$ spins on each rung span a Hilbert space $1\otimes 1=0\oplus 1\oplus 2$. The bases with total spin 0 and 2 are symmetric under $P$, they respect $A_g,B_{1g},B_{2g}, B_{3g}$ representations under group operation. The bases with total spin $S_t=1$ are antisymmetric under $P$ and respect $B_{1u},B_{2u},B_{3u}$ representations under the group operation .

$D_2\times\sigma$ has eight classes of projective representations \cite{LiuLiuWen}, the trivial projective representation is one dimensional, and the other seven nontrivial ones are two dimensional, as shown in Table \ref{projective representation} of \ref{app:Rep}. The projective representations correspond to the edge spins of the SPT phases. Due to their different edge states, these SPT phases can be completely distinguished by their different active operators, namely, by their different response to various external probing fields \cite{LiuLiuWen, spin one half}.

In the following, we give the Hamiltonians of these SPT phases (the details of calculation are given in \ref{app:H}). It has been shown in Ref.\cite{spin one half} that the $t_x, t_y, t_z$ phases can be obtained from the $t_0$ phase via onsite unitary transformations (which do not commute with the symmetry group) since their Hamiltonians and active operators can be transformed into each other by these unitary transformations. However, the remaining three SPT phases $t_{xy},\ t_{xz}, \ t_{yz}$
cannot be obtained by onsite unitary transformations from the Haldane phase $t_0$. In this sense, these three phases are more exotic comparing to the $t_x,\ t_y,\ t_z$ phases. If we collect the phases related by onsite unitary transformations into the same set, then the seven nontrivial SPT phases are separated into two disconnected sets, as shown in Fig.\ref{ladder}. In the following discussion, we denote $U_1(\theta, m)$ as a rotation of angle $\theta$ along $m$-direction for the spins on one leg, and denote $U(\theta, m)$ as a spin rotation on both legs.

\subsection{Hamiltonian for the trivial phase}
The trivial phase can be simply realized by two coupled antiferromagnetic Heisenberg chains. The Hamiltonian is given as
\begin{equation}
H_{\mathrm{Trivial}} = J_1\sum_{\tau,i}\mathbf{S}_{\tau,i}\cdot\mathbf{S}_{\tau,i+1} + J_2\sum_i\mathbf{S}_{1,i}\cdot\mathbf{S}_{2,i},
\end{equation}
where $\tau=1,2$ labels the two legs, $J_1>0$ is the intra-chain coupling and $J_2$ is the inter-chain coupling.

Above model has $SO(3)\times \sigma$ symmetry. If $J_2>0$, the system has no edge states; when $J_2<0$, the system has spin-1 edge states. In both cases, the ground state is a trivial SPT state \cite{Oshikawa, Delft}.  If the symmetry breaks down to $D_2\times \sigma$ by adding some anisotropic interactions, then generally there are no edge states no matter the inter-chain coupling is ferromagnetic or antiferromagnetic.

\subsection{Hamiltonians for the $t_0,t_x,t_y,t_z$ phases}

\begin{figure}
    \centering
    \subfigure[]{
    \includegraphics[height=20mm]{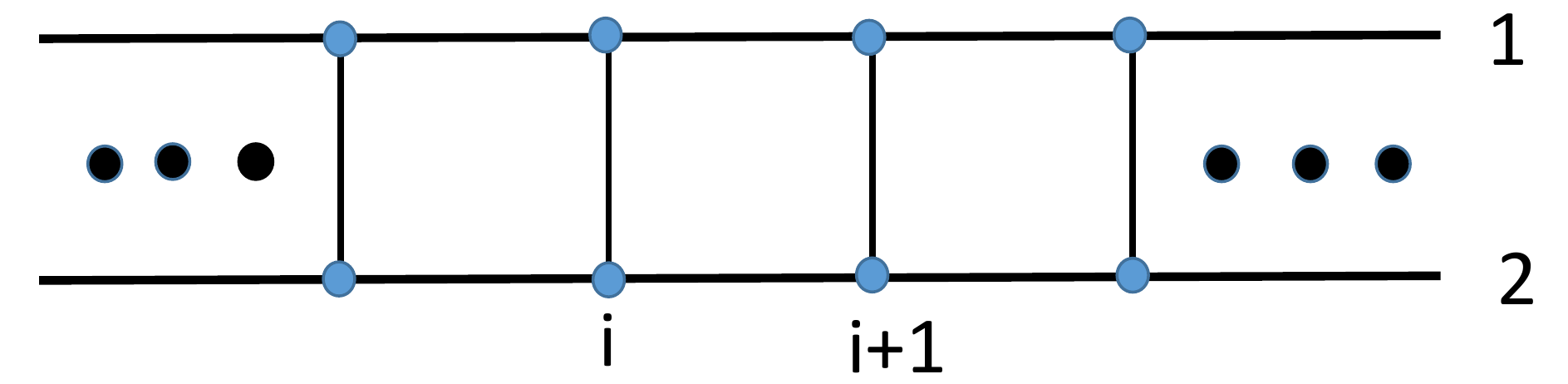}}
    \subfigure[]{
    \includegraphics[height=40mm, width=40mm]{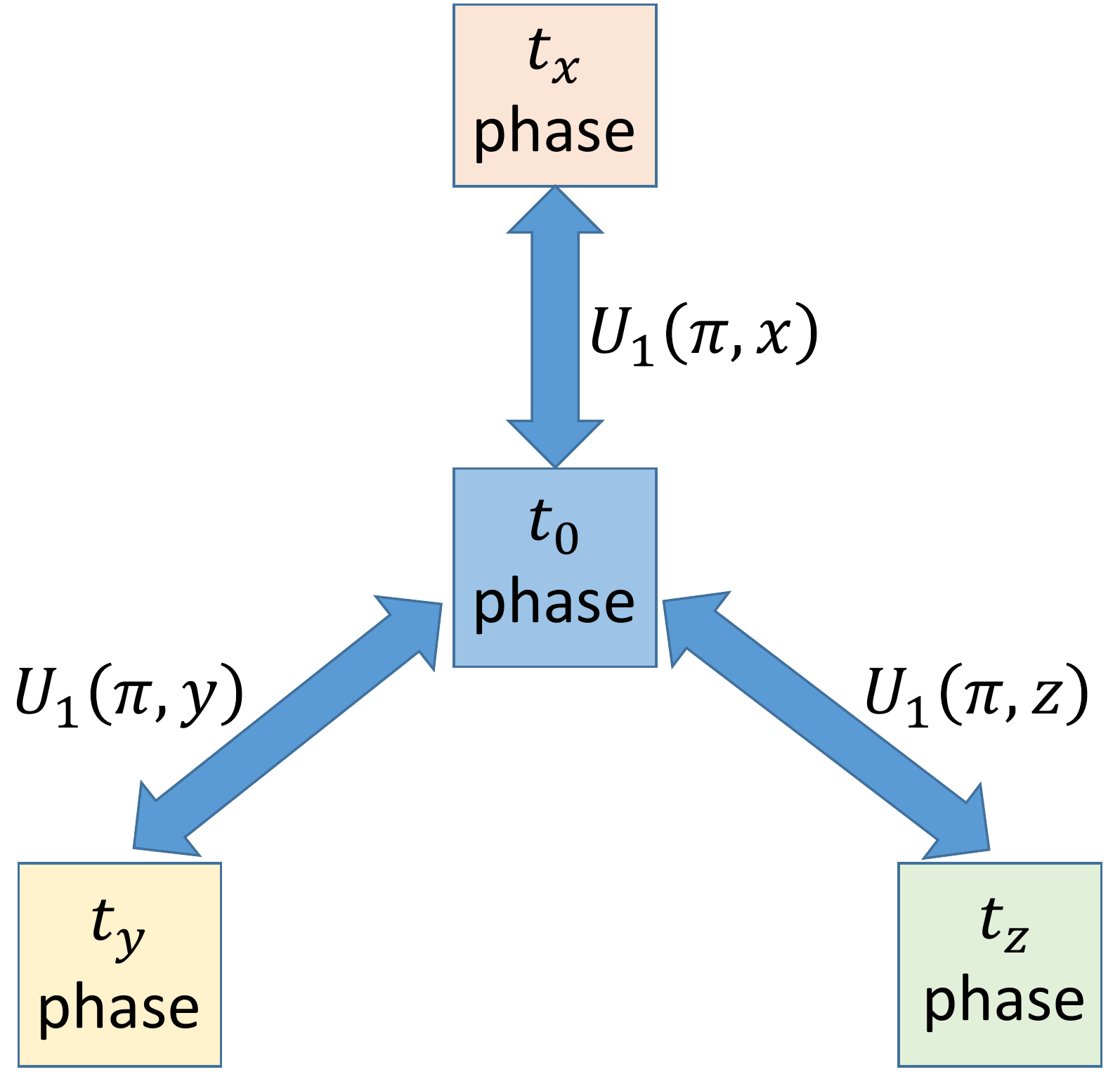}}
    \subfigure[]{
    \includegraphics[height=40mm, width=40mm]{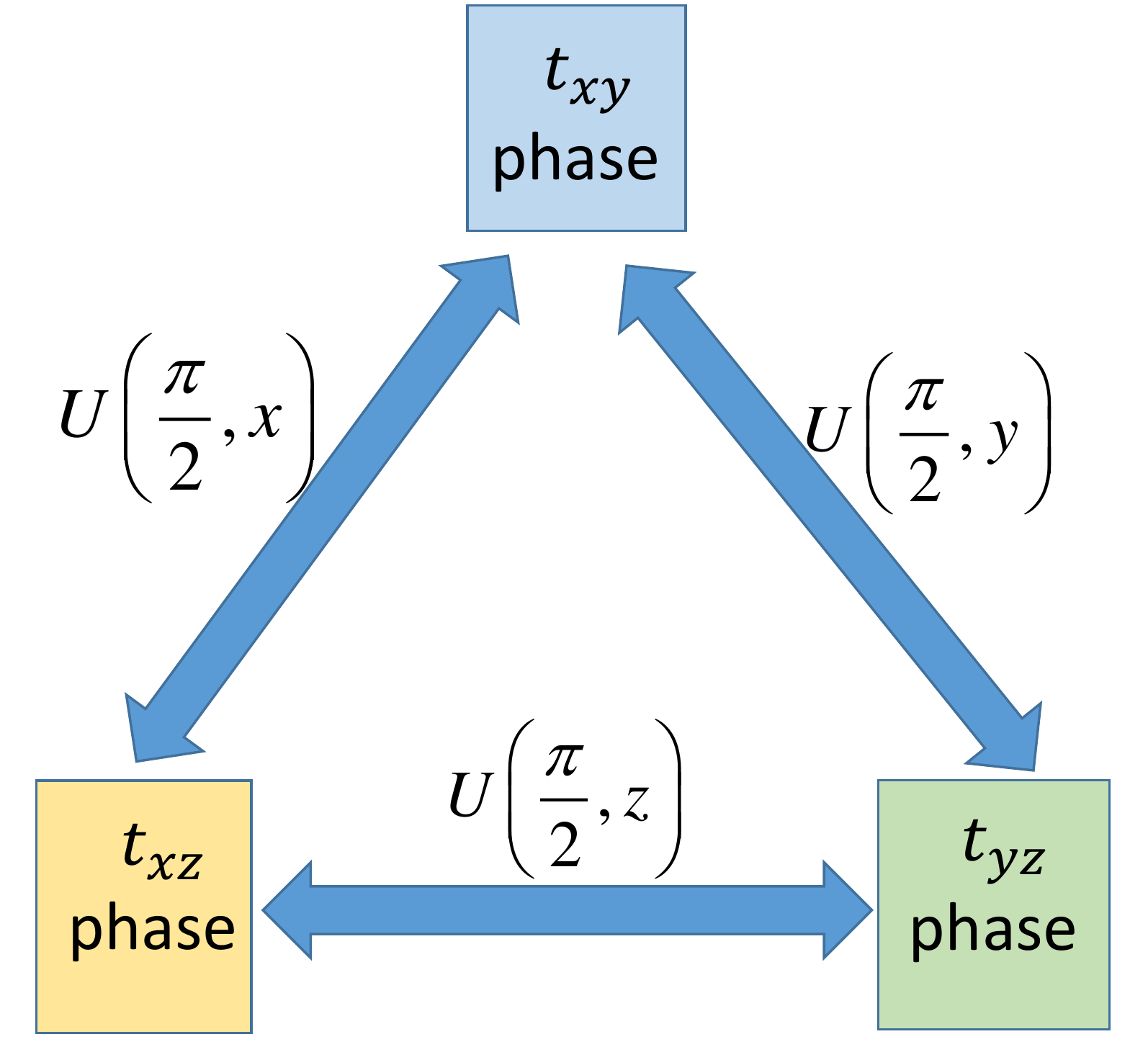}}
\caption{Schematic figure for spin-$1$ ladder system and the relations between nontrivial SPT phases. In (b), the $U_1(\pi)$ rotation only acts on one chain, while in (c) the $U(\frac{\pi}{2})$ operation acts on both chains. 
}
\label{ladder}
\end{figure}

The $t_0$ phase is the analog of the Haldane phase of spin-1 antiferromagnetic chain since they both have 2-fold degenerate edge states and their edge states respond to external magnetic field in the same way. The Hamiltonian for $t_0$ phase contains two parts,
\begin{equation}\label{T0 Hamiltonian}
H_0 = \sum_i (H^0_{i,i+1} + H^0_{i}),
\end{equation}
where $H^0_{i,i+1}$ represents rung-rung coupling between rungs $i,i+1$ and $H^0_{i}$ represents intra-rung coupling,
\begin{equation}
\begin{split}
& H^0_{i,i+1} = \sum_{\tau} J_0\left[\mathbf{S}_{\tau,i}\cdot\mathbf{S}_{\tau,i+1}
+ \beta_0 (\mathbf{S}_{\tau,i}\cdot\mathbf{S}_{\tau,i+1})^2 \right]\\
&\quad\quad\quad\quad + J_1\left[ \mathbf{S}_{\tau,i}\cdot\mathbf{S}_{\bar\tau,i+1}
+ \beta_1(\mathbf{S}_{\tau,i}\cdot\mathbf{S}_{\bar\tau,i+1})^2 \right],\\
& H^0_{i} = J_2\left[ \mathbf{S}_{1,i}\cdot\mathbf{S}_{2,i}
+ \beta_2(\mathbf{S}_{1,i}\cdot\mathbf{S}_{2,i})^2 \right],\\
\end{split}
\end{equation}
with $\tau=1,2$ and $\bar\tau = 3-\tau$ label the two legs of the ladder.  With parameters $J_0=\frac{1}{8}, \beta_0=\frac{1}{3},J_1=\frac{1}{8},\beta_1=\frac{1}{3},J_2=\frac{4}{3},\beta_2=1$, above Hamiltonian $H_0$ is frustration free and its exactly solvable ground state has spin-1/2 edge states. The $J_1$ term represents the diagonal interaction and can adiabatically be reduced to 0 without phase transition. Again, above model has an enlarged $SO(3)\times\sigma$ symmetry which can be broken down to $D_2\times\sigma$ by introducing anisotropic interactions.

The Hamiltonian  $H_x (H_y, H_z)$ of the $t_x (t_y, t_z)$  phase can be obtained from $H_0$ by spin rotation of $\pi$ angle along $x(y,z)$ direction for the spins on one of the chains.
$H_x$ also contains two parts, 
\begin{eqnarray}\label{Tx Hamiltonian}
H_x &=& \sum_i (H^x_{i,i+1} + H^x_{i}),\\
H^x_{i,i+1} &=& \sum_{\tau} J_0\left[\mathbf{S}_{\tau,i}\cdot\mathbf{S}_{\tau,i+1}
+ \beta_0 (\mathbf{S}_{\tau,i}\cdot\mathbf{S}_{\tau,i+1})^2\right]\nonumber\\
&& + J_1[ (S_{\tau,i}^xS_{\bar\tau,i+1}^x - S_{\tau,i}^yS_{\bar\tau,i+1}^y - S_{\tau,i}^zS_{\bar\tau,i+1}^z)\nonumber\\
&& +\beta_1 (S_{\tau,i}^xS_{\bar\tau,i+1}^x - S_{\tau,i}^yS_{\bar\tau,i+1}^y - S_{\tau,i}^zS_{\bar\tau,i+1}^z)^2],\nonumber\\
 H^x_{i} &=& J_2[ (S_{1,i}^xS_{2,i}^x - S_{1,i}^yS_{2,i}^y - S_{1,i}^zS_{2,i}^z)\nonumber\\
&& + \beta_2(S_{1,i}^xS_{2,i}^x - S_{1,i}^yS_{2,i}^y - S_{1,i}^zS_{2,i}^z)^2],\nonumber
\end{eqnarray}
with $S_{\tau,i}^m$  (here $m=x,y,z$) being the spin-1 operators at the $\tau$th chain and $i$th rung.
Similarly, we can obtain the Hamiltonians for $t_y$ and $t_z$ phases.

\subsection{Hamiltonians for the $t_{xy},t_{xz},t_{yz}$ phases}

The Hamiltonians for $t_{xy},t_{xz},t_{yz}$ phases cannot be obtained by simple onsite unitary transformation of $H_0$.
However, the Hamiltonians and the active operators of these three phases can be transformed into each other by onsite unitary transformations (see Fig.~\ref{ladder}).
For each phase, one exactly solvable Hamiltonian was constructed (see appendix for details). For example, the Hamiltonian for the $t_{xz}$ phase is given by
\begin{equation}\label{txz hamiltonian}
H_{xz}=\sum_i H^1_{i}+H^{2,1}_{i,i+1}+H^{2,2}_{i}+H^3_{i,i+1}+H^4_{i,i+1}
\end{equation}
where $H^1_{i}$ represents on-site term, $H^{2,1}_{i,i+1}$ represents two-body rung-rung coupling and $H^{2,2}_{i}$ represents intra-rung coupling, $H^3_{i,i+1}$ and $H^4_{i,i+1}$ represent three-body and four-body interaction. Due to its complexity, we leave the full expression of (\ref{txz hamiltonian}) to \ref{app:H}.

This Hamiltonian (\ref{txz hamiltonian}) has an enlarged symmetry. For instance, the $Z_2$ subgroup generated by $R_y$ becomes a $Z_4$ group with the generator $U(\frac{\pi}{2},y)$. (\ref{txz hamiltonian}) is also invariant under the continuous transformations $U(1)\times U(1)$, where the first $U(1)$ is generated by $\sum_i(S^{x}_{1,i})^2 - (S^{x}_{2,i})^2$ and the second $U(1)$ is generated by $\sum_i(S^{y}_{1,i})^2 - (S^{y}_{2,i})^2$.

The Hamiltonians of $t_{xy}$ and $t_{yz}$ phases can be obtained from on-site unitary transformations shown in Fig.~\ref{ladder} and will not be shown here. Contrary to the $t_0$ ($t_x, t_y, t_z$) phase, the Hamiltonian for the $t_{xz}$ ($t_{xy}$, $t_{yz}$)  phase cannot be adiabatically simplified into two-body interactions. That is to say, the three-body and four-body interactions are important to realize the $t_{xz}$  ($t_{xy}$, $t_{yz}$) SPT phase.

\begin{figure}[h]
    \centering
    \subfigure[]{
    \includegraphics[width=4cm]{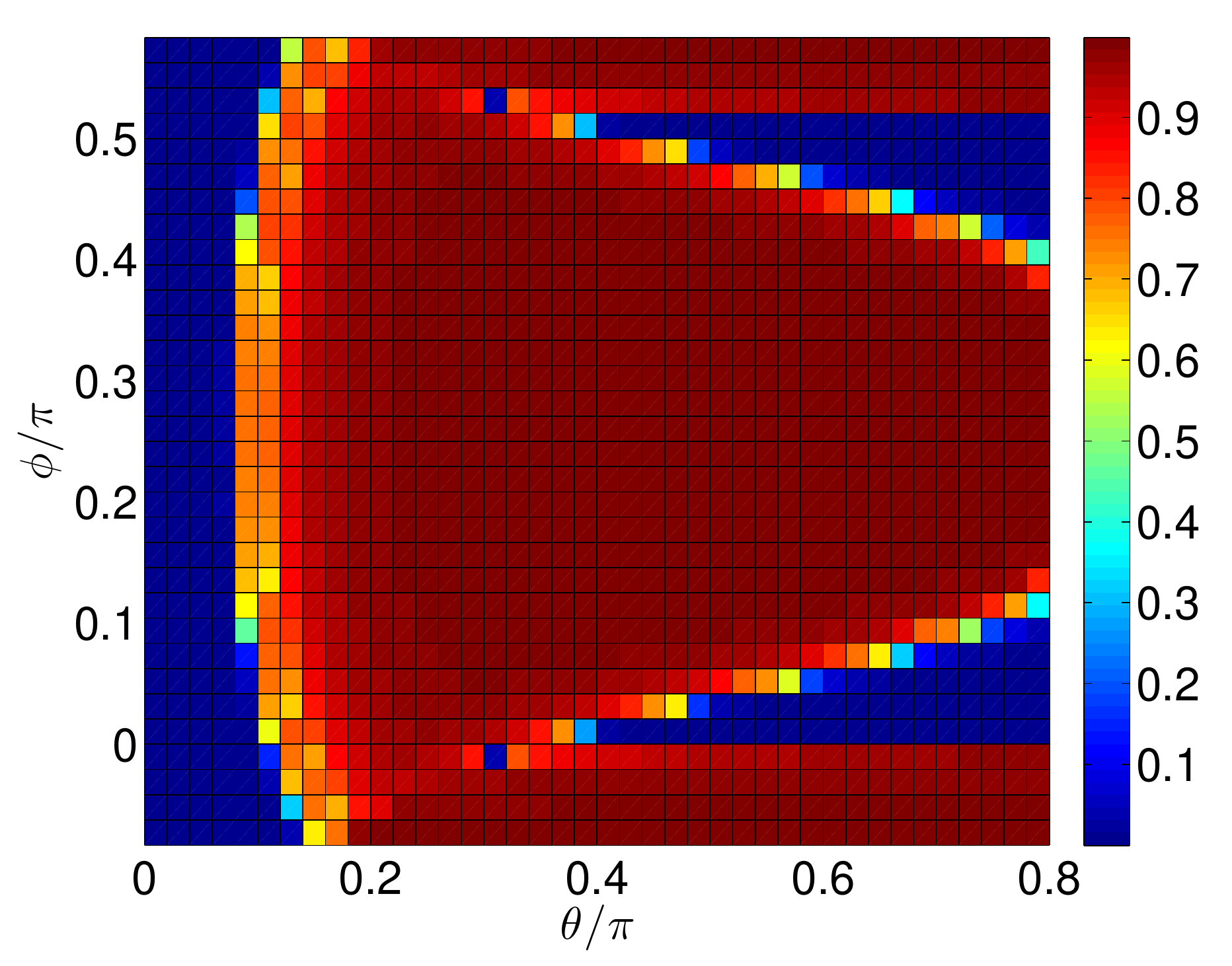}}
    \subfigure[]{
    \includegraphics[width=4cm]{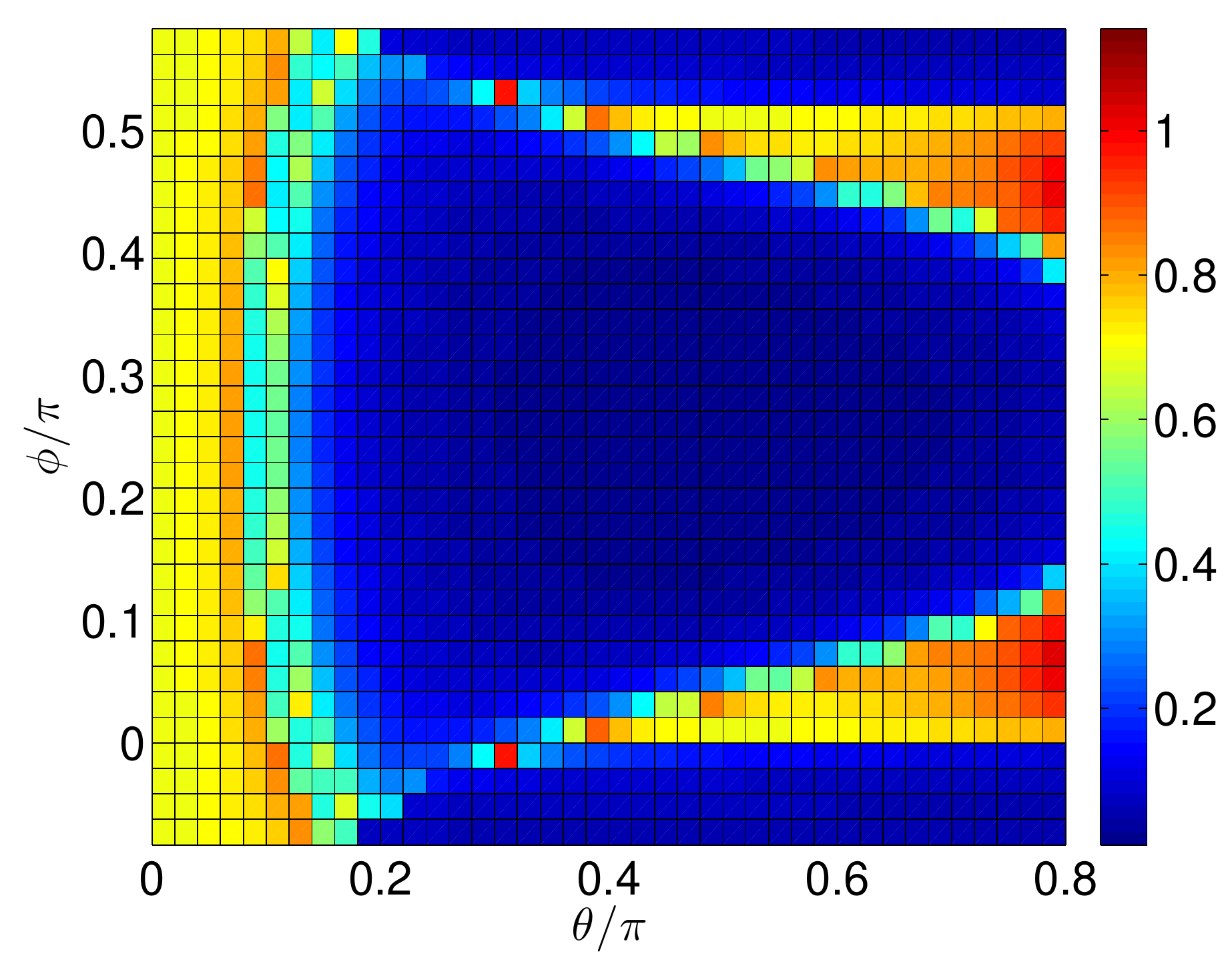}}
    \subfigure[]{
    \includegraphics[width=4cm]{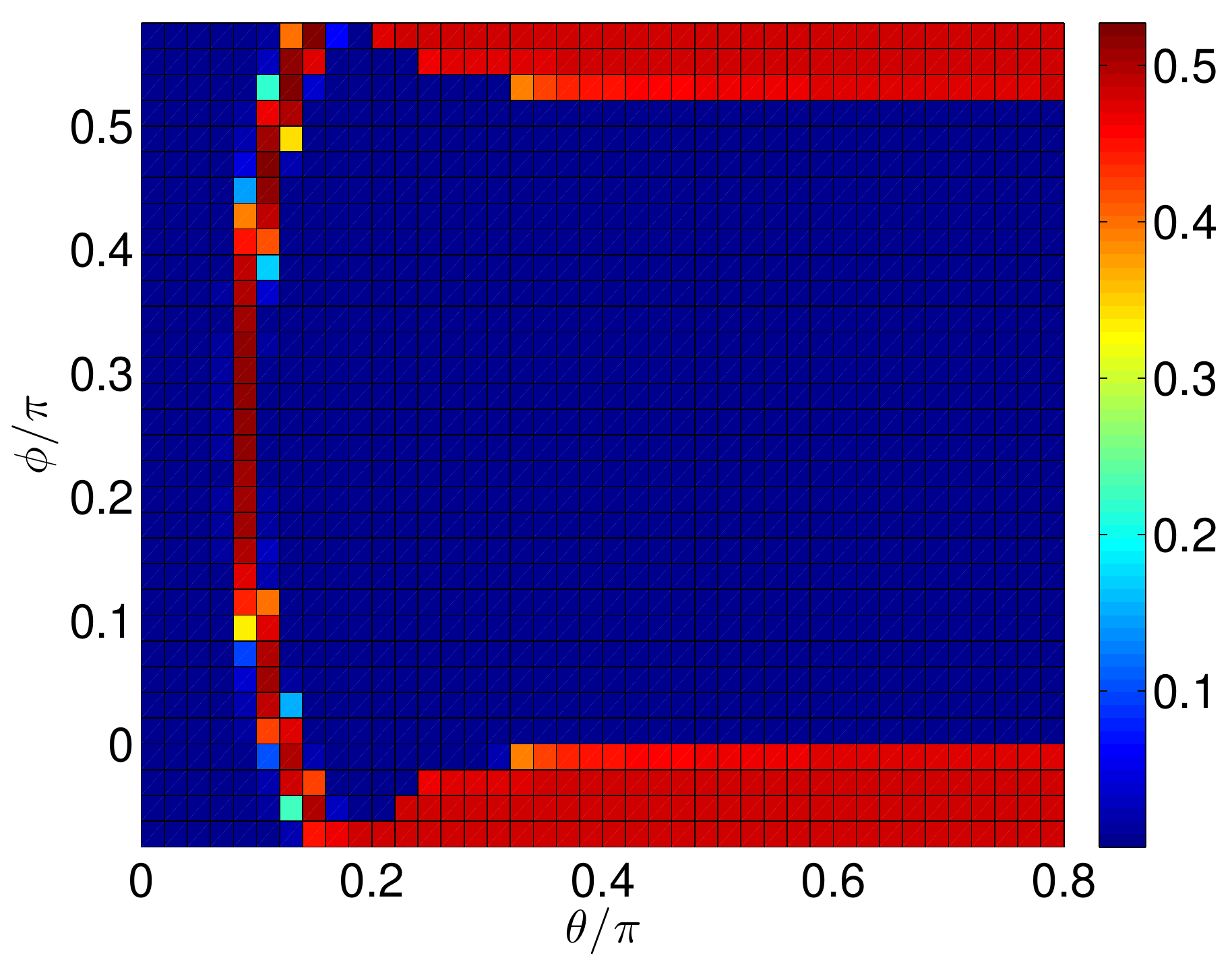}}
    \subfigure[]{
    \includegraphics[width=40mm,height=31mm]{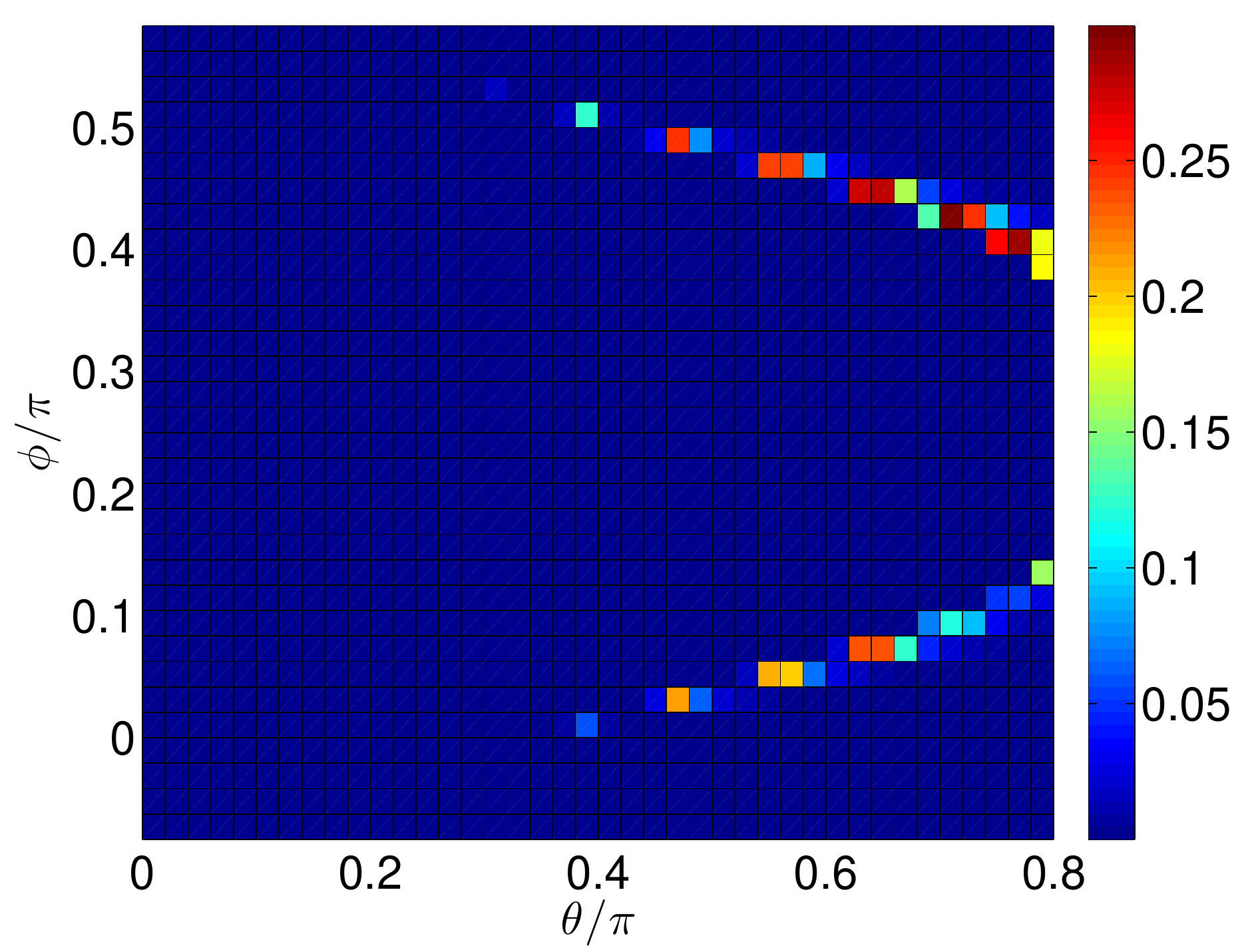}}
\caption{2D phase diagram with three SPT phases $t_{xz},t_0,t_y$. $\theta=0$ is for $H_{xz}$. $\theta=\frac{\pi}{2},\phi=0(\frac{\pi}{2})$ is for $H_0(H_y)$. (a), entanglement spectrum gap $\Delta\rho=\rho_1-\rho_2$, where $\rho_1$ and $\rho_2$ are two maximal Schmidt eigenvalues of the ladder; (b), entanglement entropy $S = -\mathrm{tr}\rho \mathrm{ln}\rho$, where $\rho$ is reduced density matrix of the ladder; (c)(d), local magnetization order $|\langle S^y_{1,i}\rangle|$ and $|\langle S^z_{1,i}\rangle|$. All the magnetization orders are in units of $\hbar$, and the same for the following figures.}
\label{2d_PD}
\end{figure}

\section{Phase transitions between SPT phases}\label{sec:trans}

In the previous section we give the Hamiltonians of the SPT phases. Here we study possible phase transitions between different SPT phases. To this end, we first consider the Hamiltonian
\begin{equation}\label{2d_PD_Ham}
H(\theta,\phi)=\mathrm{cos}\theta H_{xz} + \mathrm{sin}\theta(\mathrm{cos}\phi H_0 + \mathrm{sin}\phi H_y)
\end{equation}
which connects $t_{xz},t_0,t_y$ phases through parameters $\theta,\phi$. This model has three nontrivial SPT phases and the phase diagram can give a rough picture of phase transitions between SPT phases. The calculation is carried out using infinite time-evolving block decimation (iTEBD) method \cite{vidal} with virtual dimension $D=40$ (and the same below).

The 2D phase diagram for Hamiltonian (\ref{2d_PD_Ham}) is shown in Fig.\ref{2d_PD}. Three nontrivial SPT phases $t_{xz},t_0,t_y$ manifest themselves in Fig.\ref{2d_PD}(a), which are characterized by vanishing of entanglement spectrum gap and no magnetic orders. A trivial SPT phase locates near the center of the phase diagram, which has a large entanglement spectrum gap but no magnetic orders. The wave function for this trivial SPT phase can be adiabatically connected to a direct product state
\[
|{\rm trivial}\rangle_y=\bigotimes_i\frac{1}{2}\left(|1,0\rangle-|0,1\rangle+|0,-1\rangle-|-1,0\rangle\right)_i,
\]
which is the eigenstate of $S^{y+}_i=S^y_{1,i}+S^y_{2,i}$ with the eigenvalue 0. Most of the phase transitions seem to be second order owing to the large entanglement entropy [see Fig.\ref{2d_PD}(b)].

In above phase diagram (also see Fig.~\ref{phasediagram}), the SPT phases are separated by several symmetry breaking phases and never touch each other. That is to say, direct phase transitions DO NOT take place between any two SPT phases. The symmetry breaking phases have either $\langle S^{y\pm}\rangle$ or $\langle S^{z\pm}\rangle$ order, as shown in Fig.~\ref{2d_PD}(c) and (d) \cite{note2}.

To see generally if there exist direct phase transitions between any two of the SPT phases,  we connect the Hamiltonians $H_a, H_b$ of two SPT phases $t_a,\ t_b$ via a single parameter,
\[
H(\lambda)=\lambda H_a +(1-\lambda)H_b
\]
and study the corresponding phase diagram.  It turns out that two SPT phases may be separated by:
(A) a direct first order phase transition; (B) a trivial phase;
(C) a symmetry breaking phase;
(D) a trivial phase and symmetry breaking phases.
We will give an example for each case.

\begin{figure}
    \centering
    \subfigure[]{
    \includegraphics[width=4cm]{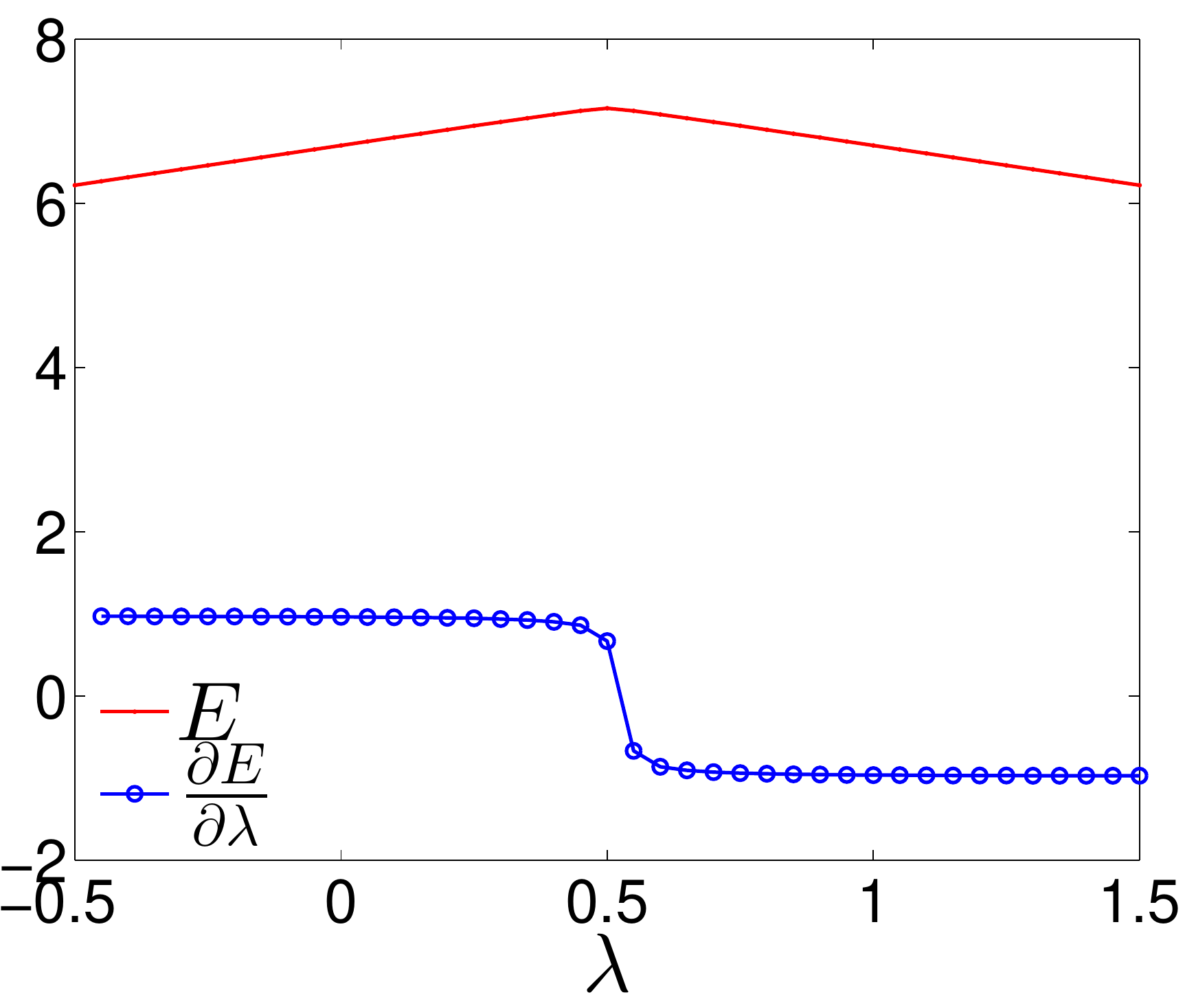}}
    \subfigure[]{
    \includegraphics[width=4cm]{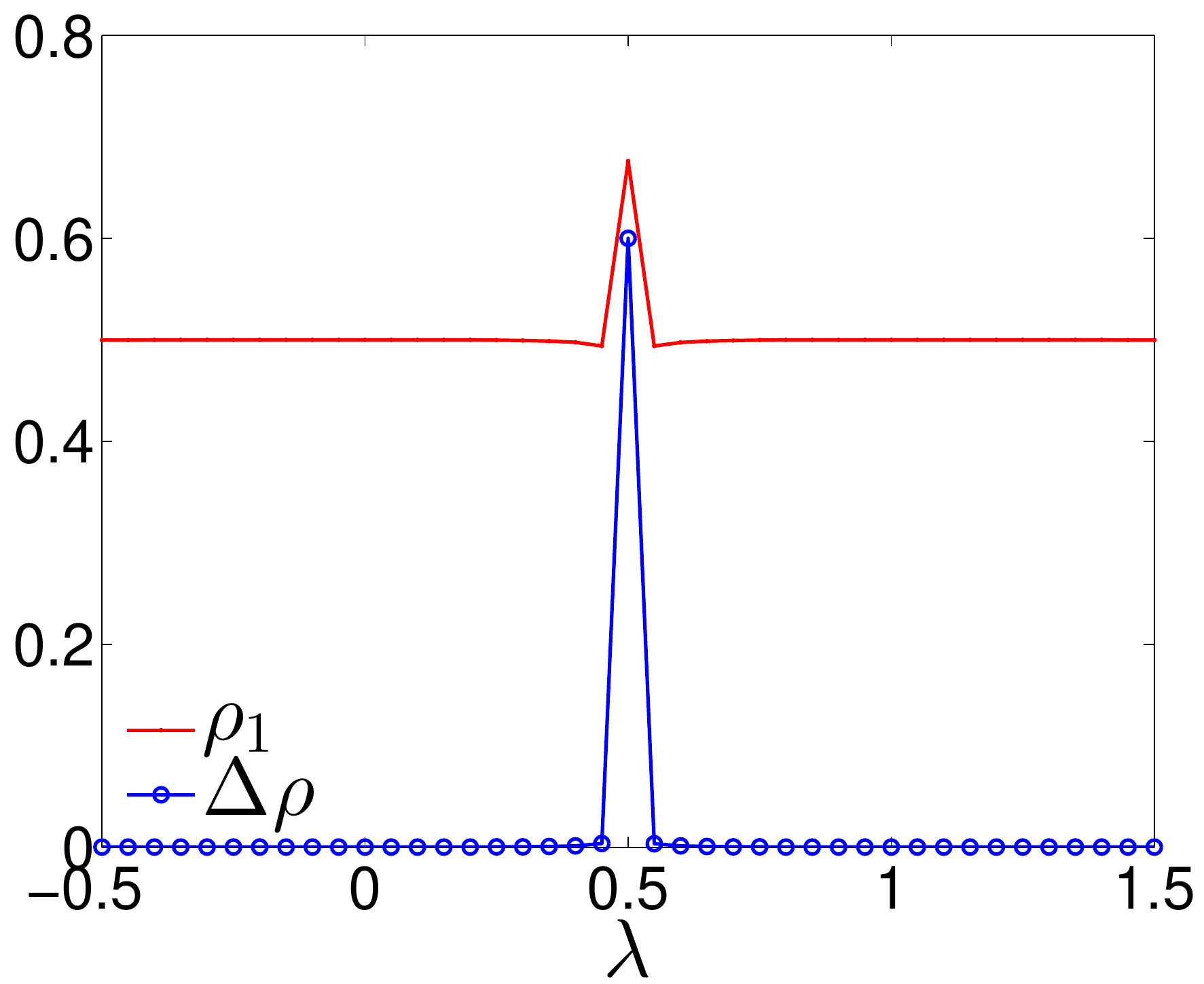}}
\caption{Phase diagram with a direct first order phase transition between $t_{xy}$ and $t_{xz}$ phases.  $\lambda=1$ is the point for $H_{xy}$, and $\lambda=0$ is for $H_{xz}$. The entanglement spectrum gap $\Delta\rho$, $\Delta\rho=0$ means the entanglement spectrum is doubly degenerate. The SPT phases are characterized by double degeneracy of entanglement spectrum and vanishing of magnetic order. }
\label{first_order}
\end{figure}

\subsection{Direct first order phase transitions}\label{caseA}
The $t_{xy}$ phase can be changed into $t_{xz}$ phase via a first order phase transition. To see this, we interpolate a Hamiltonian path between $H_{xy}$ and $H_{xz}$,
\[H =\lambda H_{xy}+(1-\lambda)H_{xz}\]
and study its phase diagram.
This model has an enlarged global symmetry $U(1)\times U(1)$, where the first $U(1)$ is generated by $\sum_i(S^{x}_{1,i})^2 - (S^{x}_{2,i})^2$ and the second $U(1)$ is generated by $\sum_i(S^{y}_{1,i})^2 - (S^{y}_{2,i})^2$.

The results are shown in Fig.\ref{first_order}.
The ground state energy $E(\lambda)$ and its derivatives are shown in Fig.\ref{first_order}(a). There is a singularity in first order derivative of the energy. Furthermore, no magnetic orders are found and the entanglement spectrum is always two-fold degenerate except the transition point [see Fig.\ref{first_order}(b)]. Above information shows that there are only two SPT phases $t_{xy}$ and $t_{xz}$, and the phase transition between them is of first order.

Direct first order phase transitions can also be found between $t_{xy}$ and $t_{yz}$ phases, or between $t_{xz}$ and $t_{yz}$ phases, or between any two of $t_x,t_y,t_z$ phases.

\subsection{Phase transition through a trivial phase}\label{caseB}
Between $t_0$ and $t_x$ phases we did not find a direct phase transition in the phase diagram of the model
\[
H=\lambda H_{0} + (1-\lambda)H_x.
\]
This model has an enlarged (accidental) global $U(1)$ symmetry generated by $\sum_iS^{x+}_i$.

\begin{figure}
    \centering
    \subfigure[]{
    \includegraphics[width=4cm]{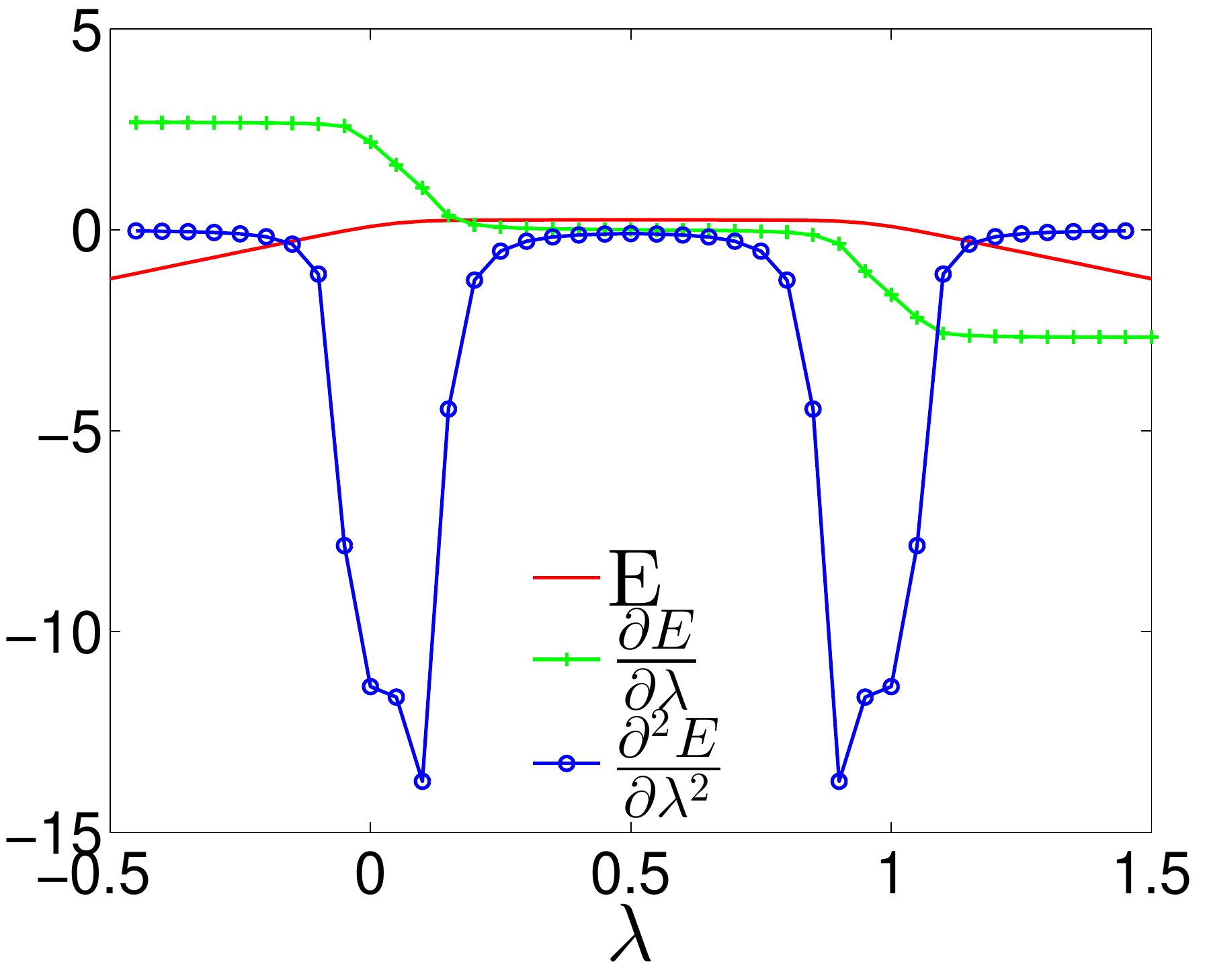}}
    \subfigure[]{
    \includegraphics[width=4cm]{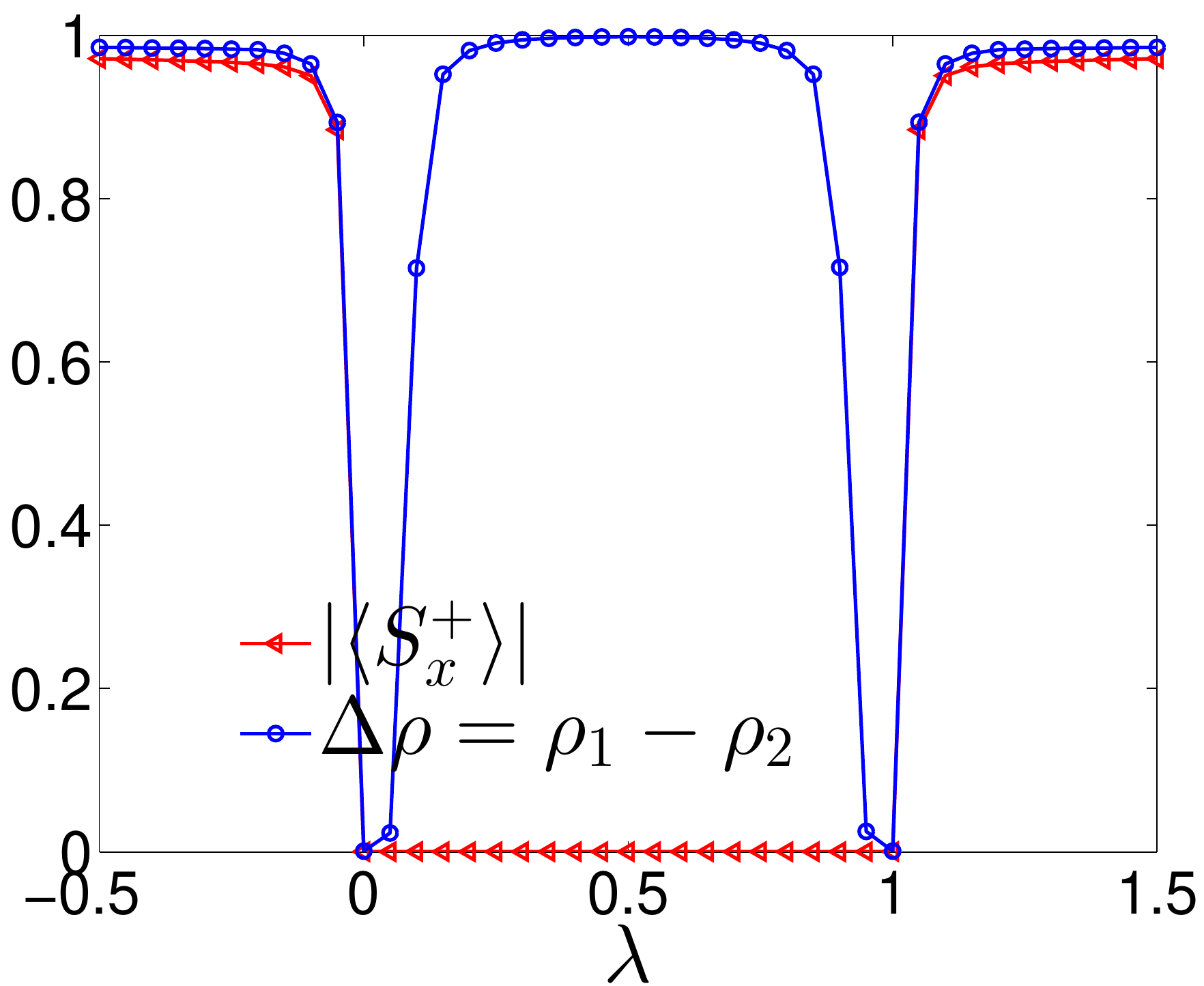}}
\caption{Phase transitions interpolating $t_0$ and $t_x$. (a), energy and its derivatives; (b), the local order is $|\langle S_{i}^{x+} \rangle| = (-1)^i\langle S_{1,i}^x+S_{2,i}^x \rangle$. Between two SPT phases there is a trivial phase characterized by a large entanglement spectrum gap $\Delta\rho$ and vanishing magnetic orders.
}
\label{trivial}
\end{figure}

It turns out that the two phases $t_0$ and $t_x$ are separated by a trivial SPT phase, as shown in Fig.~\ref{trivial}. The trivial phase can be adiabatically connected to the direct product state
\[
|{\rm trivial}\rangle_x=\bigotimes_i\frac{1}{2}\left(|1,0\rangle - |0,1\rangle - |0,-1\rangle + |-1,0\rangle\right)_i.
\]
At each rung above state is the eigenstate of $S^{x+}_i=S_{1,i}^x+S_{2,i}^x$ with the eigenvalue 0 \cite{note1}.

The phase diagram in Fig.\ref{trivial} shows that both $t_0$ and $t_x$ phases are narrow regions sandwiched by a trivial phase and a stripe Neel ordered phase with nonvanishing $|\langle S^{x+}_i\rangle|$. If the accidental $U(1)$ symmetry is removed by some perturbations, then an extra symmetry breaking phase will appear to separate the trivial phase and each SPT phase(see Sec.\ref{caseC}). The physical reason behind this will be discussed in Sec. \ref{discussPT}.

\subsection{Phase transition through a symmetry breaking phase}\label{caseC}
\begin{figure}
    \centering
    \subfigure[]{
    \includegraphics[width=4cm]{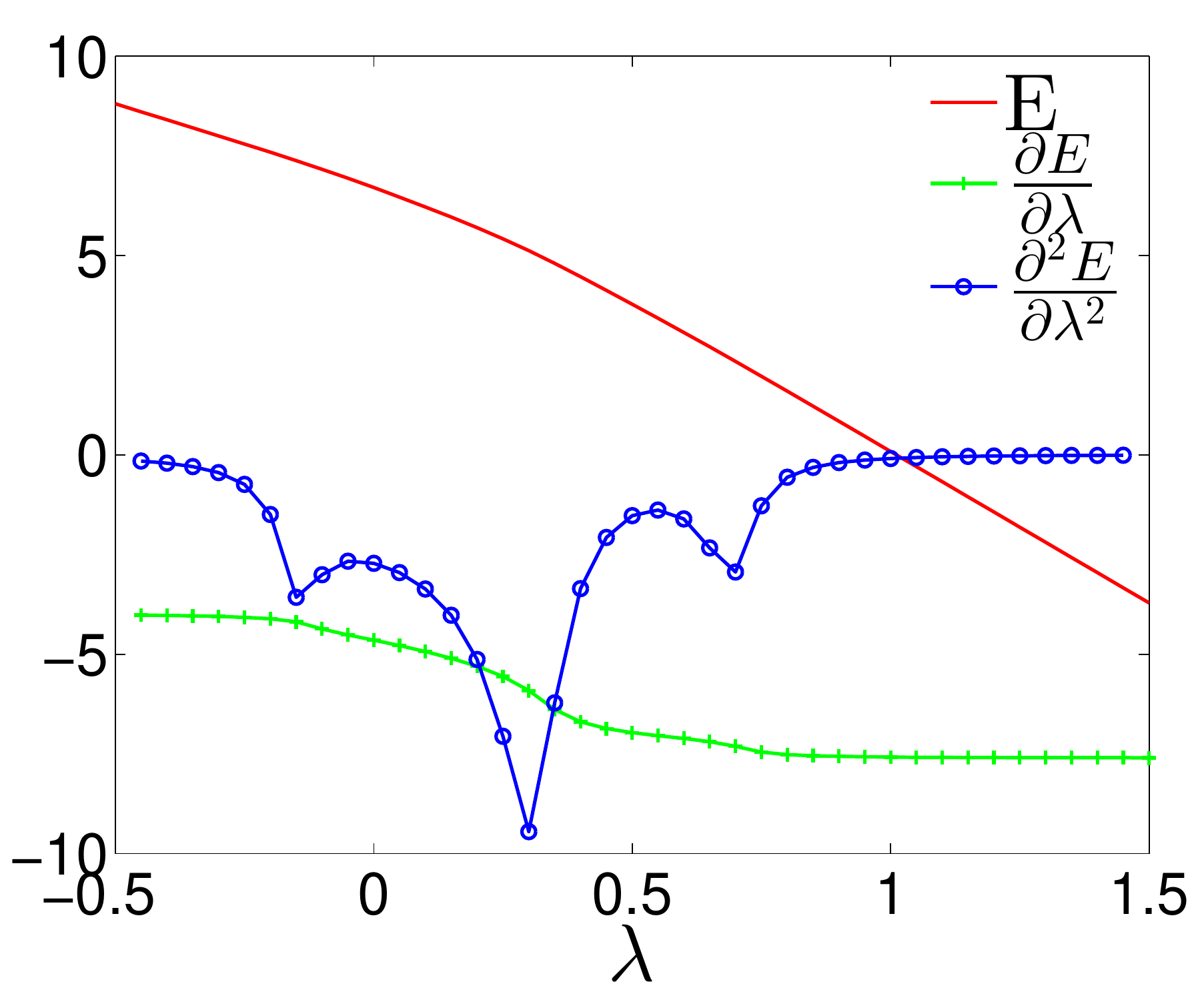}}
    \subfigure[]{
    \includegraphics[width=4cm]{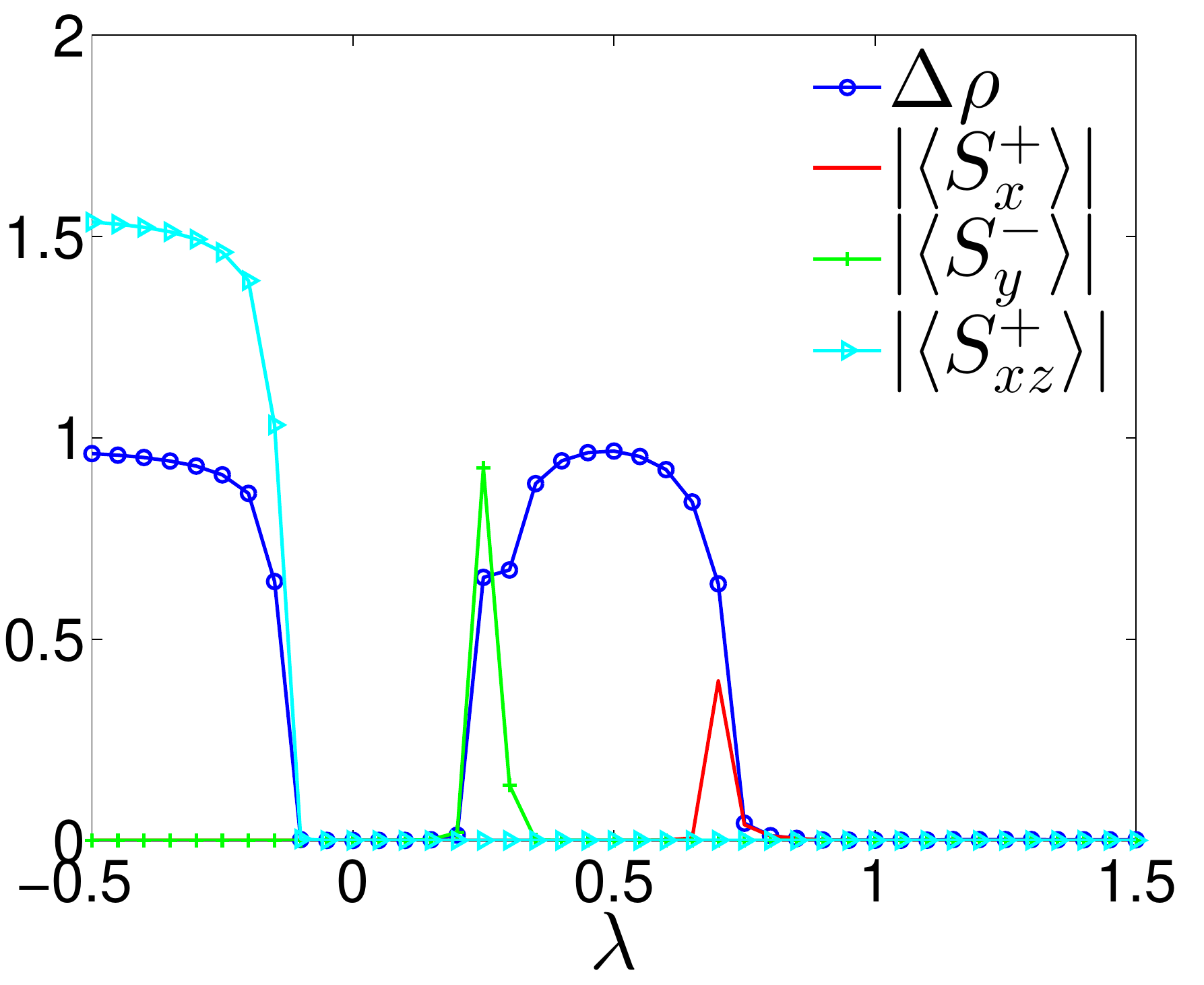}}
\caption{Phase diagram for $t_0$ and $t_{xz}$ phase transition through a trivial phase and two symmetry breaking phases.
Between the trivial SPT phase and $t_{xz}$ phase, there is an antiferromagnetic ordered phase where the order is $|\langle S^{y-}_i\rangle|=(-1)^i\langle S^y_{1,i} - S^y_{2,i}\rangle$. While between the trivial SPT phase and $t_0$ phase, an antiferromagnetic ordered phase exists in which the order is $|\langle S^{x+}_i\rangle| = (-1)^i\langle S^x_{1,i}+S^x_{2,i}\rangle$. The region $\lambda<-0.1$ is a new symmetry breaking phase with order $|\langle S^{xz,+}\rangle| =(-1)^i\langle S^{xz}_{1,i}+S^{xz}_{2,i}\rangle $.}
\label{symmetry_breaking}
\end{figure}

In studying the model
\[
H = \lambda H_{0} + (1-\lambda)H_{xz},
\]
we find different symmetry breaking phases separating the trivial SPT phase and $t_0 (t_{xz})$ phases as shown in Fig.\ref{symmetry_breaking}.

In Fig.\ref{symmetry_breaking}, there is a small $S^{z+}$ order coexisting with the $S^{x+}$ order at the same parameter region with $|\langle S^{z+}_i\rangle| = (-1)^{i+1}\langle S^z_{1,i}+S^z_{2,i}\rangle$. This order is relatively small with the largest value $1.4\times10^{-2}$, so we did not show it in the figure.

The trivial phase in Fig.\ref{symmetry_breaking} can be adiabatically connected to the direct product state
\[
|{\rm trivial}\rangle_y=\bigotimes_i\frac{1}{2}\left(|1,0\rangle-|0,1\rangle+|0,-1\rangle-|-1,0\rangle\right)_i.
\]
At each rung above state is the eigenstate of $S^{y+}_i=S^y_{1,i}+S^y_{2,i}$ with the eigenvalue 0.

The phase diagram in Fig.~\ref{symmetry_breaking} tells us two pieces of information: (1) the non-trivial SPT phases are sandwiched by symmetry breaking phases; (2) two SPT phases (including trivial SPT phase) are separated by a symmetry breaking phase.

Now we study another example,
\[
H = \lambda H_x + (1-\lambda)H_{xz}
\]
which connects the $t_x$ and $t_{xz}$ phases.

\begin{figure}
    \centering
    \includegraphics[width=8cm,height=6cm]{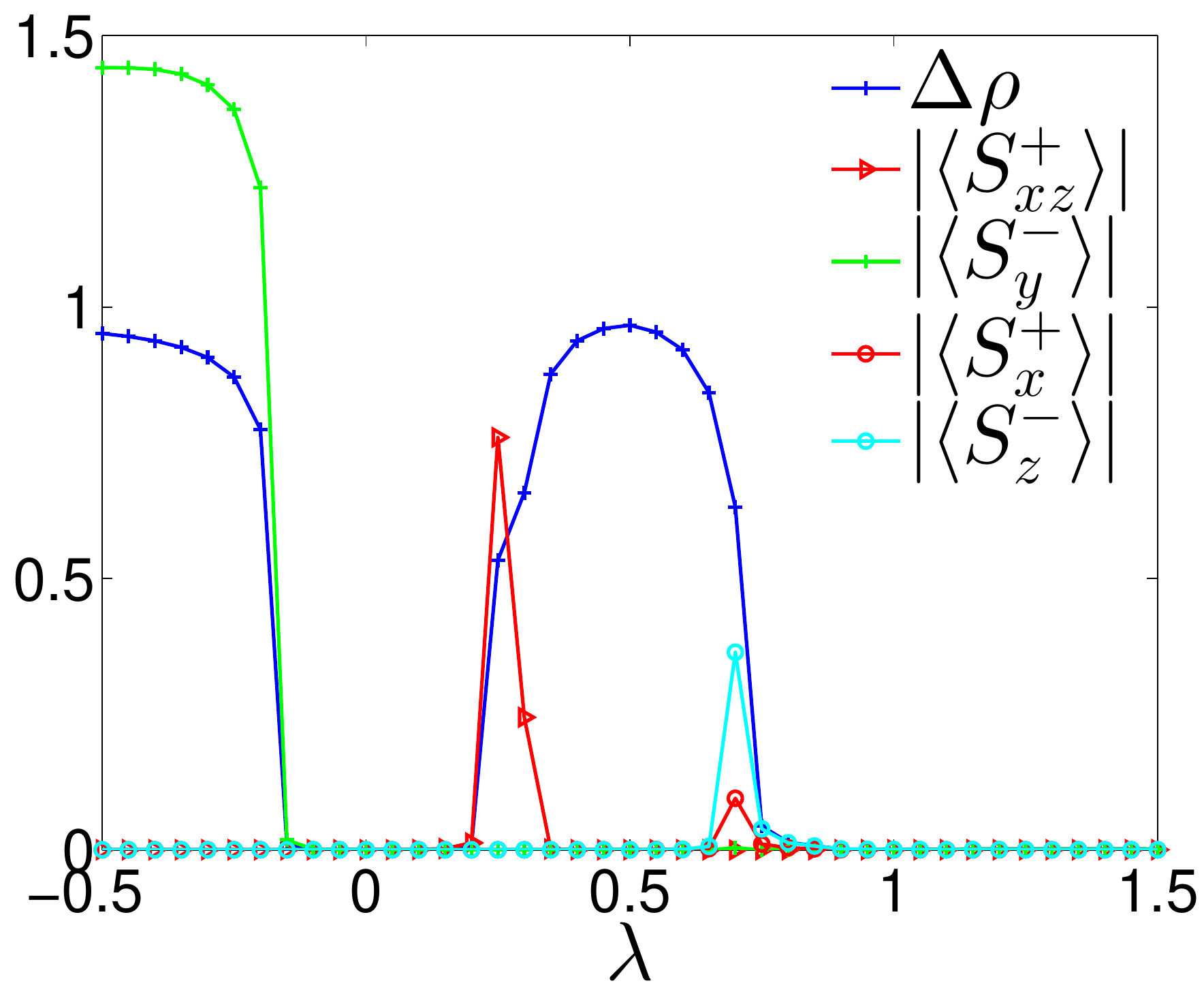}
\caption{Phase diagram for $t_x$ and $t_{xz}$ phase transition through two symmetry breaking phases and a trivial phase. Between the two nontrivial SPT phases, there is a trivial phase and two antiferromagnetic ordered phases. The magnetic orders are $|\langle S^{xz,+}_i\rangle|= (-1)^{i}\langle S^{xz}_{1,i}+S^{xz}_{2,i}\rangle$, $|\langle S^{y-}\rangle|=(-1)^i\langle S^{y}_{1,i}-S^y_{2,i}\rangle$, $|\langle S^{x+}_i\rangle|= (-1)^{i}\langle S^x_{1,i}+S^x_{2,i}\rangle$, $|\langle S^{z-}_i\rangle|= (-1)^{i+1}\langle S^z_{1,i}-S^z_{2,i}\rangle$, where the $|\langle S^{x+}_i\rangle|$ order is smaller than
$|\langle S^{z-}_i\rangle|$ order.
 When $\lambda\leq-0.15$, the antiferromagnetic ordered phase with the magnetic order $|\langle S^{y-}\rangle|$ appears.}
\label{trivial_SB}
\end{figure}

The phase diagram is very similar to the previous one. As shown in Fig.~\ref{trivial_SB}, the intermediate trivial phase can be adiabatically connected to the direct product state
\[
|{\rm trivial}\rangle_{y-}=\bigotimes_i\frac{1}{2}\left(|1,0\rangle+|0,1\rangle-|0,-1\rangle-|-1,0\rangle\right)_i.
\]
At each rung above state is the eigenstate of $S^{y-}_i=S^y_{1,i}-S^y_{2,i}$ with the eigenvalue 0.
The sharp peak of order parameter $|\langle S^{z-}_i\rangle|$ ($|\langle S^{xz,+}_i\rangle|$) indicates that there is a symmetry breaking phase separating the trivial SPT phase and $t_{x}$ $(t_{xz})$ phase.

Once again, the phase diagram in Fig.~\ref{trivial_SB} shows that generally two SPT phases are separated by a symmetry breaking phase. Similar results can also be obtained between $t_x$ and $t_y$ phases, or between $t_y$ and $t_{xz}$ phases.

\subsection{Summary of phase transitions between SPT phases}\label{sumPT}

According to the numerical results we obtained,  two different SPT phases are separated either by a first order phase transition or by intermediate trivial or/and symmetry breaking phases. Direct second order phase transitions between nontrivial SPT phases are never observed.

Above results are reasonable for discrete symmetry group protected topological phases. Notice that nontrivial SPT phases are disordered by strong quantum fluctuations (also see Sec. \ref{discussPT}), and the quantum fluctuations are resultant from competition between different classical orders. Since discrete symmetry can be spontaneously broken in 1D, SPT phases are always sandwiched by symmetry breaking phases if the symmetry is not enhanced \cite{spin one half}. This explains why nontrivial SPT phases are always separated by one or several ordered phase. A direct first order transition between two SPT phases is possible but is not interesting. Our results suggest that in 1D generally different nontrivial SPT phases protected by discrete symmetry DO NOT have direct continuous phase transitions if the symmetry is not enhanced to a continuous one. This is an important message of the present paper.

However, if the symmetry group which is broken in the intermediate ordered phase is enhanced to a continuous one, then the intermediate ordered phase will vanish in 1D owning to Mermin-Wagner theorem \cite{Mermin}. In that case, the two SPT phases have direct second order transition. That is to say, the continuous phase transition between different SPT phases (including the trivial SPT phase) is `protected' by continuous symmetry. This phenomenon has been observed in previous numerical studies \cite{Zheng-Cheng Gu, pollmann 1, spin one half, HHTu}.

\section{Effective field theory of SPT phases}\label{discussPT}

In the following we will interpret some results of previous sections from low energy effective field theory. We will start from a single spin-1 chain model, discuss possible phases and phase transitions, then we couple two chains to form ladder models \cite{Bosonization}.

\subsection{Nonlinear Sigma Model description of a single chain}
{\bf $SO(3)$ symmetry}. For a single spin-1 chain Heisenberg model $H=\sum_i\mathbf S_i\cdot\mathbf S_{i+1}$, it was shown that the low energy effective field theory is the topological $SO(3)$ nonlinear sigma model (NLSM) \cite{Haldane}
\begin{eqnarray}
&&Z=\int D\pmb n(x,\tau) e^{-S_{\rm d} - S_{\rm \theta}},\nonumber\\
&&S_{\rm d}=\iint d\tau dx {1\over g}(\partial_\mu \pmb n)^2, \nonumber\\
&&S_{\rm \theta}=i\theta\iint d\tau dx{1\over 4\pi}\pmb n\cdot(\partial_\tau\pmb n\times\partial_x\pmb n).
\end{eqnarray}
where $\theta=2\pi S$ with $S=1$. Denoting $|\pmb n\rangle$ as the spin-1 coherent state, then we can define Berry connection  $A_\mu=\langle\pmb n|\partial_\mu |\pmb n\rangle$ and the Berry curvature $F=\partial_\tau A_x-\partial_x A_\tau$. The $\theta$-term can also be written as $S_{\rm \theta}=i{\theta\over 4\pi}\iint d\tau dx F$. 


The dynamic term $S_{\rm d}$ is unimportant since under renormalization group (RG) the coupling constant $g$ flows to infinity. The $\theta$-term is very important, from which we can read off the ground state wave function under periodic boundary condition (here we ignore the dynamic term),
\begin{eqnarray}\label{eq:Haldane}
|\psi\rangle = \int D\pmb n(x) e^{i{\theta\over4\pi}\oint A_xdx}\bigotimes_x|\pmb n(x)\rangle.
\end{eqnarray}
Under open boundary condition (supposing the temporal boundary condition is periodic), the $\theta$-term
\begin{eqnarray*}
S_{\rm \theta}&=&i{\theta\over 4\pi}\iint d\tau dx F \\
&=& {i\over2}\left(\oint d\tau\cdot \pmb A|_{x=0}-\oint d\tau \cdot\pmb A|_{x=L}\right)
\end{eqnarray*}
effectively behaves like Berry phase terms of spin-1/2 particle living on each boundary, which explains the existence of spin-1/2 edge states of the Haldane phase \cite{TK94}.

{\bf $O(2)$ symmetry}. Now if we add an anisotropy term $\sum_i D(S^z_{i})^2$ to the Hamiltonian, then the symmetry of the chain becomes $O(2)=U(1)\times Z_2$, where $U(1)$ is continuous spin rotation along $\hat z$-direction and $Z_2$ is generated by a spin rotation of $\pi$ along $\hat x$-direction. In the low energy effective theory, the dynamic term becomes
\[
S_{\rm d}=\iint d\tau dx {1\over g}(\partial_\mu \pmb n)^2 + D  n_z^2.
\]
In the large $D\to + \infty$ limit, the vector $\pmb n$ lies in the $xy$-plain and the integrand of the $\theta$-term vanishes. The resultant ground state wave function is a trivial product state
\[
 |\psi\rangle=\int D\pmb n(x)\bigotimes_x |\pmb n\rangle_{n_z=0}=\bigotimes_x\int d\pmb n |\pmb n\rangle_{n_z=0}=\bigotimes_i|0\rangle_i,
\]
where $|0\rangle$ is the 0-component of spin-1.

With the decreasing of the value of $D$, the $z$-component of $\pmb n$ increases. After a critical point $D_c$ the $\theta$-term plays an important role and the system enters the Haldane phase. If we ignore the dynamic term, then the ground state is given by (\ref{eq:Haldane}) under periodic boundary condition. At the transition point $D_c$, the fluctuation in the $\theta$-term is so strong such that the bulk gap closes.

When $D$ further decreases, $\hat z$ becomes easy axes as $D<0$ and at some critical $D_c'$ the systems form antiferromagnetic order $|\langle S_z\rangle|$, which spontaneously breaks the $Z_2$ symmetry. So the Haldane phase is sandwiched by a trivial phase and a symmetry breaking phase.

{\bf $Z_2\times Z_2$ symmetry}. Now we further introduce anisotropy into the Heisenberg interaction, then the dynamic term becomes
\[
S_{\rm d}=\iint d\tau dx {1\over g}[J_x(\partial_\mu n_x)^2+(\partial_\mu n_y)^2 +(\partial_\mu n_z)^2] + D n_z^2,
\]
and the symmetry of the chain reduces into $Z_2\times Z_2$. We will see that the Haldane phase is wrapped by three symmetry breaking phases.

In the large $J_x$ limit, the systems falls in a symmetry breaking phase $M_x$ and has an antiferromagnetic order $|\langle S_x\rangle|$.  In the large negative $D$ limit, the system is in $M_z$ phase which contains antiferromagnetic order $|\langle S_z\rangle|$. On the other hand, if $D$ is positive but not too large (otherwise the system enters the trivial phase), then the vector $\pmb n$ favors $xy$-plane, at the same time if $J_x<1$, then the vector $\pmb n$ favors $yz$-plane. Resultantly within a certain region of $D$ and $J_x$, the system will fall in the $M_y$ phase and has antiferromagnetic order $|\langle S_y\rangle|$.

The Haldane phase locates near the center of the three ordered phases, where all the three order parameters are fluctuating such that $\pmb n$ can be pointing to arbitrary directions. Thus the $\theta$-term is non-vanishing and gives rise to a SPT phase. In other words, the Haldane phase is a consequence of the competition between the three orders.

Especially, the previous direct transition from the large $D$ trivial phase to the Haldane phase is split into two phase transitions, since there is an intermediate symmetry breaking phase $M_y$ (if $J_x<1$) or $M_x$ (if $J_x>1$). Owing to the decreasing of the symmetry, the anisotropy terms $D$ and $J_x$ suppress the topological $\theta$-term and trigger the symmetry breaking, which prevents the touching between the trivial phase and the Haldane phase except at the high symmetry point $J_x=1$.

\subsection{Nonlinear Sigma Model description for ladder SPT phases}
{\bf $t_0$ phase}. Now we consider two spin-1 Heisenberg chains coupled with inter-chain interaction $H_i^0=J[4-(\mathbf S_{1,i}\times\mathbf S_{2,i})^2]=J[(\mathbf S_{1,i}\cdot \mathbf S_{2,i}) + (\mathbf S_{1,i}\cdot \mathbf S_{2,i})^2]$. Now the system has $SO(3)$ symmetry and the effective field theory is given by
\begin{eqnarray}
Z&=&\int D\pmb n(x,\tau) \exp(-S_{\rm d} - S_{\rm \theta}),\label{eq:t0}\\
S_{\rm d}&=&\iint d\tau dx {1\over g}[(\partial_\mu \pmb n_1)^2 +(\partial_\mu \pmb n_2)^2] -J(\pmb n_1\times \pmb n_2)^2, \nonumber\\
S_{\rm \theta}&=&i\theta\iint d\tau dx{1\over 4\pi}[\pmb n_1\cdot(\partial_\tau\pmb n_1\times\partial_x\pmb n_1) \nonumber\\
&&\ \ \ \ \ \ \ \ \ \ \ \ \ \ \ \ \ \ \ \ \ \ \   + \pmb n_2\cdot(\partial_\tau\pmb n_2\times\partial_x\pmb n_2)].\nonumber
\end{eqnarray}
where $\theta=2\pi$.

In the limit $J\to\infty$, the interchain term behaves as a projector such that at each rung $\pmb n_1$ and $\pmb n_2$  are orthogonal to each other, namely $\pmb n_1(x,\tau)\cdot\pmb n_2(x,\tau)=0$. Under this condition, the topological term is equal to
\[
S_\theta = i\theta'\iint d\tau dt {1\over4\pi}\pmb m\cdot(\partial_\tau\pmb m\times \partial_x\pmb m)
\]
where $\pmb m={1\over \sqrt2}(\pmb n_1 +\pmb n_2)$ is a unit vector. Notice that $\theta' =\sqrt2\theta=2\sqrt2\pi$ is not quantized. However, since $2\pi<\theta'<3\pi$, under RG $\theta'$ will flow to its fixed point $2\pi$ \cite{RGflow1, RGflow2, RGflow3}. As a result, the strongly coupled spin-1 ladder is similar to a single Haldane chain and has spin-1/2 edge states ({\it i.e.} the $t_0$ phase).

{\bf Trivial phase}. Now suppose $J$ is small such that $\pmb n_1$ and $\pmb n_2$ are almost independent. Then we can treat the interchain coupling $J$ term as a perturbation. Owing to the $\theta$-term for both $\pmb n_1$ and $\pmb n_2$, under open boundary condition, each chain has a spin-1/2 Berry phase term at the end. The interchain coupling $-J(\pmb n_1\times \pmb n_2)^2\sim J(\pmb n_1\cdot\pmb n_2)^2$ will cause antiferromagnetic correlation between the edge states of the two chains, as a result the Berry phase terms tend to cancel each other such that effectively there is no net Berry phase at the boundary(which is equivalent to $\theta=0$). In other words, the two spin-1/2 edge spins of the two chains form a singlet and finally the system has no edge state. So the system is in a trivial phase.

When increasing $J$, effectively the $\theta$ varies from 0 to $2\pi$ (here the $\theta$ is the fixed point value after RG flow), so there will be a continuous phase transition (at $\theta=\pi$) from the trivial phase to the Haldane phase.

{\bf $t_x$ phase}. The $t_x$ phase differs from the $t_0$ phase by the interchain coupling. Since the Hamiltonians of the two phases are related by an onsite transformation $U_1(\pi,x)$, the effective field theory for the $t_x$ phase can be obtained from (\ref{eq:t0}) by replacing $\pmb n_1$ with $\tilde {\pmb n}_1=(n_1^x, -n_1^y, -n_1^z)^T$, namely, replacing $-J(\pmb n_1\times \pmb n_2)^2$ with $-\tilde J(\tilde{\pmb  n}_1\times \pmb n_2)^2$ and keeping the rest part unchanged. 

Similar to the previous discussion, it can be shown that the $t_x$ phase with large $\tilde J$ also has two-fold degenerate edge states. On the other hand, small $\tilde J$ results in a trivial phase (if the system has no translational symmetry, this trivial phase is the same phase with the preceding one). Increasing $\tilde J$ to a critical value, the system will undergo a continuous phase transition from the trivial phase to the $t_x$ phase.

{\bf Phase transitions from $t_0$ to $t_x$}. Above we have illustrated a possible path from the $t_0$ phase to the $t_x$ phase:  
\[
t_0 \to \mathrm{trivial }\to t_x
\] 
by tuning $J$ to zero then turning on $\tilde J$. Similar result can be obtained if the interchain coupling is parameterized by $\lambda$
\[
-[\lambda J(\pmb n_1\times \pmb n_2)^2 +(1-\lambda)\tilde J(\tilde{\pmb n}_1\times \pmb n_2)^2].
\]
$\lambda=0$ gives the $t_x$ phase and $\lambda=1$ is the $t_0$ phase.  In the intermediate region near $\lambda=1/2$, the interchain coupling behaves like a projection operator locking $\pmb n_1$ and $\pmb n_2$ into a scalar, so the resultant phase is trivial.

Until now the system has $U(1)$ symmetry, so direct phase transitions take place between the nontrivial SPT phases and the trivial SPT phase. Even though, different nontrivial SPT phases  are untouched. If we lower the symmetry by introducing anisotropic Heisenberg interaction, then similar to the discussion for a single chain, even the direct phase transition between nontrivial SPT phase and the trivial phase will be prohibited, and intermediate symmetry breaking phases will occur.

In principle, the effective theory for the $t_{xz}, t_{xy}, t_{yz}$ phases can be obtained from their microscopic Hamiltonians. However, since the interactions contain quadrupole terms and many body interactions, the effective theories are not the same as usual topological NLSMs. We will not discuss about the details here.

In above discussion, the symmetry operations acting on the fields $\pmb n_1, \pmb n_2$ are the same for all SPT phases, as defined by the symmetry group. On the other hand, if we allow the symmetry operations to act differently for different SPT phases, then the effective field theory for all the SPT phases can be written in the same form, {\it i.e.} the $SO(3)$ topological NLSM \cite{CXu}. If we discretize space and time, the effective field theory of SPT phases can also be described by topological NLSMs with group cocycles \cite{wen}.

\section{Conclusion}\label{sec:sum}
In summary, we have studied SPT phases protected by symmetry group $D_2\times\sigma$. We realized all the SPT phases on spin-1 ladder models and studied the phase transitions between different SPT phases with iTEBD method. We did not find continuous phase transitions between any two SPT phases. Instead, two SPT phases are separated by either a first order transition or by one (or several) intermediate phase. We interpreted part of our results via topological nonlinear sigma model effective field theory. We further conjecture that in 1D  generally continuous phase transition cannot take place between SPT phases protected by discrete symmetry group, unless the symmetry is enhanced to a  continuous one, or a continuous symmetry emerges near the critical point in the low energy and long wavelength limit. If the critical point between two SPT phases exists, then the local order parameters corresponding to discrete symmetry breaking have power-law correlations, these quasi-long-range orders can be easily turned into long-range orders (unless the order is associated with continuous symmetry breaking) by perturbations preserving the discrete symmetry. Our results provide some hints for studying the transitions between SPT phases in higher dimensions \cite{DHLee2015,LuLee2014,ChenLee2012}.

We do not rule out the possibility that the Hamiltonian only has discrete symmetries but at very low energy and small momentum a continuous symmetry emerges. That is to say, in some special fine tuned models, direct second order phase transitions may exist between SPT phases protected by discrete symmetries owning to the emergent continuous symmetry at the critical point \cite{DHLee2015}. However,  if symmetry preserving perturbations are added to the Hamiltonian, the direct continuous transition should be unstable and may split into several transitions between which discrete symmetry breaking phase appears. 

\section{Acknowledgement}\label{sec:ack}
We thank Li You, Hong-Hao Tu, R. Orus, Peng Ye, Zheng-Cheng Gu and Chao-Ming Jian for helpful discussions. JYC is supported by the MOST 2013CB922004 of the National Key Basic Research Program of China, and by NSFC (Nos. 91121005 and 11374176). ZXL is supported by NSFC 11204149 and Tsinghua University Initiative Scientific Research Program. We acknowledge the computation time on the clusters in IASTU.

\appendix

\section{linear and projective representations of the symmetry group}\label{app:Rep}

In this appendix, we give the linear and projective representations of the symmetry group $D_2\times \sigma$, as shown in Tab.~\ref{linear representation} and Tab.~\ref{projective representation}, respectively.

\begin{table}[b]
\caption{Linear representation of $D_2\times\sigma$, $P$ is the interchain permutation operator. Notice the notation $O^{\pm} = O_1\pm O_2$ is a single-rung operator, where $O_1$ is the operator for the first chain and $O_2$ is that for the second chain.}
\label{linear representation}
\begin{tabular}{l|c|c|c|c|c}
\hline\hline
\qquad & \multicolumn{3}{|c|}{$R_z$ $R_x$ $P$}& bases & operators\\
\hline
$A_g$  & \multicolumn{3}{|c|}{1 \quad 1\quad 1} & $|0,0\rangle$,$|2,0\rangle$, $\frac{1}{\sqrt{2}}(|2,2\rangle + |2,-2\rangle)$ & $S_x^{2+}$, $S_y^{2+}$, $S_z^{2+}$\\
$B_{1g}$ & \multicolumn{3}{|c|}{1\quad -1 \quad 1} & $\frac{1}{\sqrt{2}}(|2,2\rangle - |2,-2\rangle)$ & $S_z^+$\\
$B_{2g}$ & \multicolumn{3}{|c|}{-1\quad -1 \quad 1} & $\frac{1}{\sqrt{2}}(|2,1\rangle - |2,-1\rangle)$ & $S_y^+$\\
$B_{3g}$ & \multicolumn{3}{|c|}{-1\quad 1\quad 1} & $\frac{1}{\sqrt{2}}(|2,1\rangle + |2,-1\rangle)$ & $S_x^+$\\
\hline
$A_u$ & \multicolumn{3}{|c|}{1\quad 1\quad -1} & \qquad & $S_x^{2-}$, $S_y^{2-}$, $S_z^{2-}$\\
$B_{1u}$ & \multicolumn{3}{|c|}{1\quad -1\quad -1} & $|1,0\rangle$ & $S_z^-$\\
$B_{2u}$ & \multicolumn{3}{|c|}{-1\quad -1\quad -1} & $\frac{1}{\sqrt{2}}(|1,1\rangle + |1,-1\rangle)$ & $S_y^-$\\
$B_{3u}$ & \multicolumn{3}{|c|}{-1\quad 1\quad -1} & $\frac{1}{\sqrt{2}}(|1,1\rangle - |1,-1\rangle)$ & $S_x^-$\\
\hline\hline
\end{tabular}
\end{table}

\begin{table}[t]
\caption{Projective representations and the corresponding SPT phases of $D_2\times\sigma$. The effective operators for the active operators are $(\sigma_x, \sigma_y, \sigma_z)$. The active operators can split the degeneracy of ground states on open boundaries.}
\label{projective representation}
\begin{tabular}{l|c|c|c|c|c}
\hline\hline
\qquad & \multicolumn{3}{|c|}{ $R_z$\qquad\qquad $R_x$ \qquad\qquad $P$} & Active operators & SPT phases\\
\hline
$E_0$ & \multicolumn{3}{|c|}{ $1$ \qquad\qquad $1$ \qquad\qquad $1$} & \qquad & I (trivial)\\ 
\hline
$E_1$ & \multicolumn{3}{|c|}{ I \qquad\qquad $i\sigma_z$ \qquad\qquad $\sigma_y$} & ($S_{x,y,z}^{2-}$,$S_z^-$,$S_z^+$) & $t_{xy}$\\
\hline
$E_3$ & \multicolumn{3}{|c|}{ $\sigma_z$ \qquad\qquad I \qquad\qquad $i\sigma_y$} & ($S_{x,y,z}^{2-}$,$S_x^-$,$S_x^+$) & $t_{yz}$\\
\hline
$E_5$ & \multicolumn{3}{|c|}{ $i\sigma_z$ \qquad\qquad $\sigma_x$ \qquad\qquad I} & ($S_x^+$, $S_y^+$, $S_z^+$) & $t_0$\\
\hline
$E_7$ & \multicolumn{3}{|c|}{ $\sigma_z$ \qquad\qquad $i\sigma_z$ \qquad\qquad $i\sigma_x$} & ($S_{x,y,z}^{2-}$,$S_y^-$,$S_y^+$) & $t_{xz}$\\
\hline
$E_9$ & \multicolumn{3}{|c|}{ $i\sigma_z$ \qquad\qquad $\sigma_x$ \qquad\qquad $i\sigma_x$} & ($S_x^+$, $S_y^-$, $S_z^-$) & $t_{x}$\\
\hline
$E_{11}$ & \multicolumn{3}{|c|}{ $i\sigma_z$ \qquad\qquad $i\sigma_x$ \qquad\qquad $\sigma_z$} & ($S_x^-$, $S_y^-$, $S_z^+$) & $t_{z}$\\
\hline
$E_{13}$ & \multicolumn{3}{|c|}{ $i\sigma_z$ \qquad\qquad $i\sigma_x$ \qquad\qquad $i\sigma_y$} & ($S_x^-$, $S_y^+$, $S_z^-$) & $t_{y}$\\
\hline\hline
\end{tabular}
\end{table}

\
\section{constructing the hamiltonian for SPT phase}\label{app:H}
In this appendix, we show the method by which we obtain the Hamiltonian for each SPT phase. The method is closely related to the way for obtaining Affleck-Kennedy-Lieb-Tasaki (AKLT) Hamiltonian \cite{AKLT}. Here we follow the method used in Ref.\cite{spin one half}. The procedures contain three steps:(i) construct a matrix product state (MPS) wave function with given edge states which are described by the projective representations; (ii) construct the parent Hamiltonian for the MPS using projection operators; (iii) simplify the parent Hamiltonian by adiabatic deformations. The linear representations have been shown in Table \ref{linear representation}, and the projective representations in Table \ref{projective representation}. The Hilbert space of the direct product of two projective representations can be reduced to a direct sum of linear representations, which has already been listed in Ref.\cite{LiuLiuWen}. In the following, we will illustrate the method using the Hamiltonian of $t_0$ phase as an example.

\begin{figure}
    \centering
    \subfigure[]{
    \includegraphics[width=4cm]{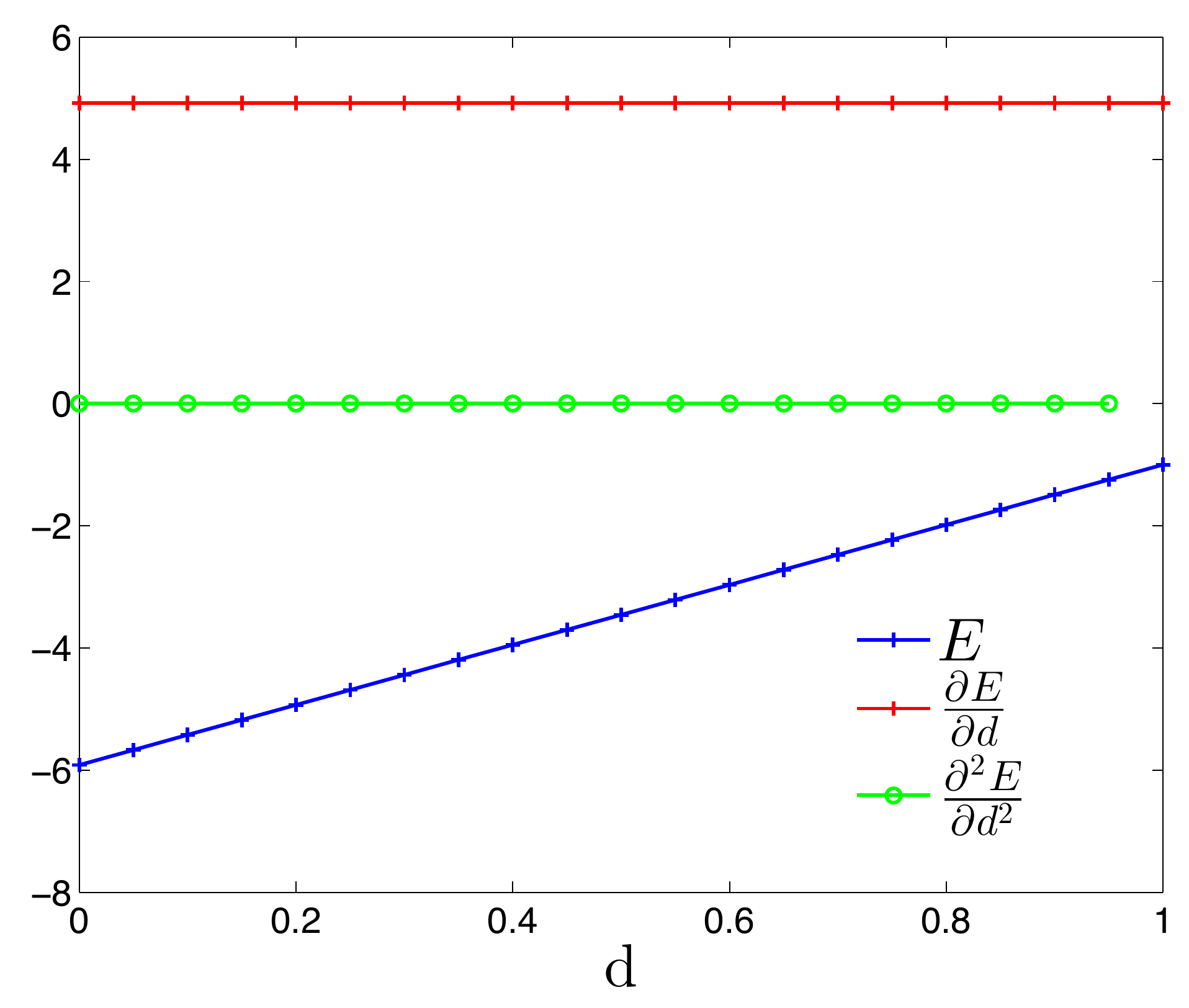}}
    \subfigure[]{
    \includegraphics[width=4cm]{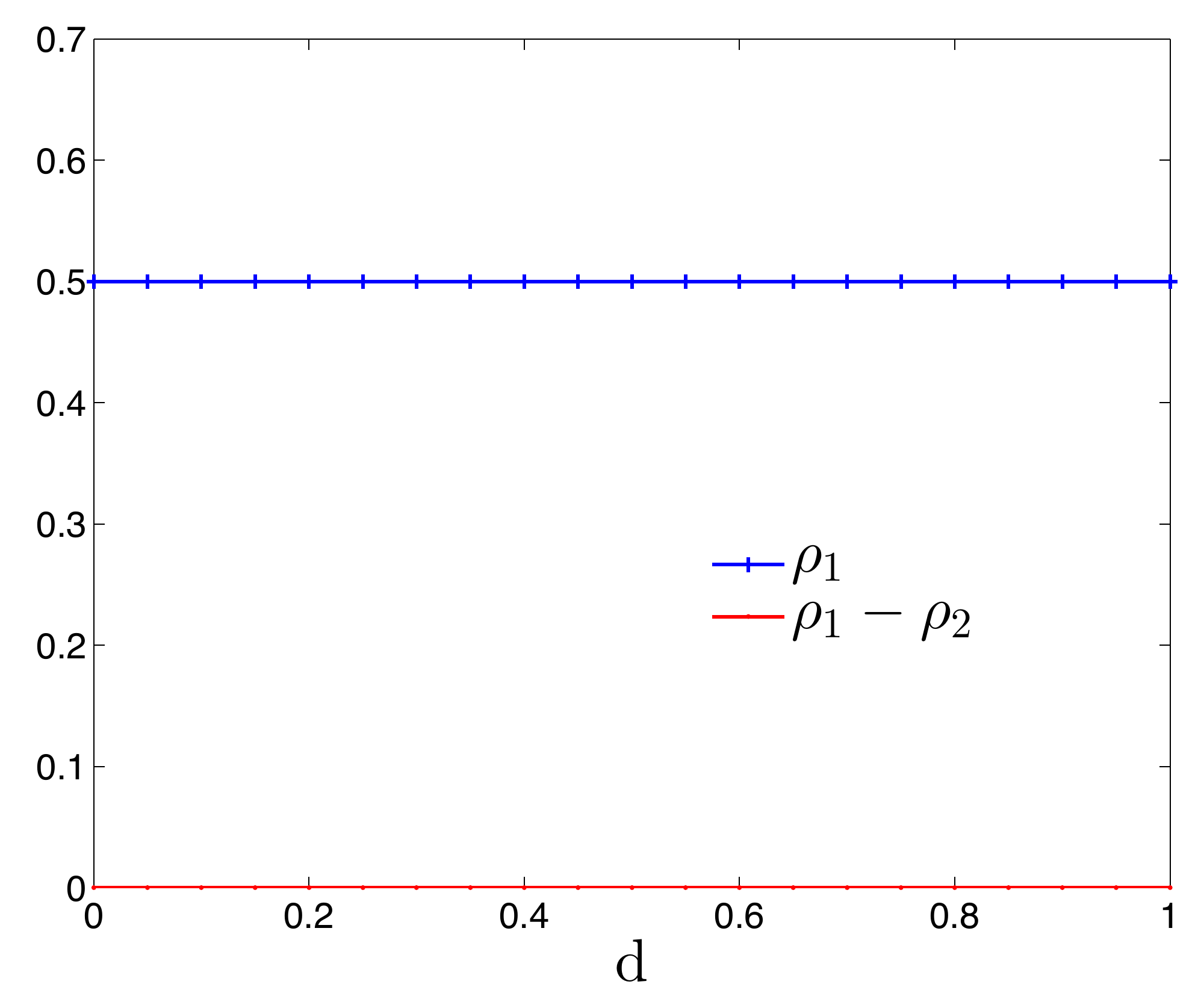}}
\caption{Deform the exact Hamiltonian of $t_0$ phase continuously with a parameter $d$. When $d=1$, it is the exact Hamiltonian.
When $d=0$, we have dropped the three-body and four-body interaction terms. (a) Energy, first energy derivative and second
energy derivative. (b) Largest of Schmidt eigenvalue and entanglement spectrum gap. The numerical method is iTEBD with virtual dimension $D=40$.}
\label{E5E6}
\end{figure}

For $t_0$ phase, the edge states can be described by $E_5$ and $E_6$ representations. ($E_6$ is not shown explicitly in Table \ref{projective representation}. Both $E_6$ and $E_8$ can be found in Ref.\cite{LiuLiuWen}.) In an ideal MPS, every rung is represented by a direct product of two projective representations, which can be reduced to four linear representations. Here we choose $E_5\otimes E_6$ with the Clebsch-Gordan (CG) coefficients:
\begin{equation}
\begin{split}
& E_5\otimes E_6=A_u\oplus B_{1u}\oplus B_{2u}\oplus B_{3u}\\
& C^{A_u}=\sigma_x, C^{B_{1u}}=i\sigma_y, C^{B_{2u}}=\sigma_z, C^{B_{3u}}=I;\\
\end{split}
\end{equation}
where the CG coefficients have been chosen real. From the CG coefficients, we can write an ideal MPS which is invariant (up to a phase) under the symmetry group $D_2\times\sigma$:
\begin{equation}
|\psi\rangle=\sum_{\{m_1,...,m_N\}}Tr(A^{m_1}...A^{m_N})|m_1...m_N\rangle
\end{equation}
with $A^m=e^{i\theta_m}BC^m$. Here $B$ is the CG coefficient of decomposing the product representation $E_5\otimes E_6$ into one-dimensional representation which was chosen to be $C^{A_u}$ and $e^{i\theta_m}$ can be absorbed into the spin bases. Thus we have the four bases on each rung:
\begin{equation}
\begin{split}
& |\phi_1\rangle=-\frac{1}{\sqrt{2}}(|1,-1\rangle-|-1,1\rangle);\quad |\phi_2\rangle=|1,0\rangle-|0,1\rangle;\\
& |\phi_3\rangle=-|0,-1\rangle+|-1,0\rangle; \quad |\phi_4\rangle=\frac{1}{\sqrt{2}}(|1,-1\rangle-|-1,1\rangle).
\end{split}
\end{equation}
The orthonormal bases for the MPS state of two neighboring rungs $i, i+1$ are:
\begin{equation}
\begin{split}
& |\psi_1\rangle=\frac{1}{\sqrt{5}}(|\phi_1\rangle_i\otimes|\phi_1\rangle_{i+1} + |\phi_3\rangle_i\otimes|\phi_2\rangle_{i+1});\\
& |\psi_2\rangle=\frac{1}{2}(|\phi_1\rangle_i\otimes|\phi_3\rangle_{i+1}-|\phi_3\rangle_i\otimes|\phi_1\rangle_{i+1})\\
& |\psi_3\rangle=\frac{1}{2}(|\phi_2\rangle_i\otimes|\phi_1\rangle_{i+1}-|\phi_1\rangle_i\otimes|\phi_2\rangle_{i+1})\\
& |\psi_4\rangle=\sqrt{\frac{5}{24}}(|\phi_2\rangle_i\otimes|\phi_3\rangle_{i+1}+\frac{4}{5}|\phi_1\rangle_i\otimes|\phi_1\rangle_{i+1}
-\frac{1}{5}|\phi_3\rangle_i\otimes|\phi_2\rangle_{i+1}).
\end{split}
\end{equation}
The parent Hamiltonian $H_{0,ex}=\sum_iH_i=-\sum_i\sum_{a=1}^4|\psi_a\rangle\langle\psi_a|$ has many terms,
$H_i=H_2 + d\times(H_3 + H_4)$, where $d=1$, $H_3$ and $H_4$ represent three-body and four-body interaction, respectively.
When we tune the parameter $d$ from 1 to 0, there is no phase transition, shown in Fig.\ref{E5E6}. Further dropping the constant leads to the Hamiltonian in Eq.(\ref{T0 Hamiltonian}).

The same method can be used for $t_{xz}$ phase, where we choose projective representation $E_7\otimes E_8$ as our starting point,
\begin{equation}
\begin{split}
& E_7\otimes E_8 = A_g\oplus B_{2g}\oplus A_u\oplus B_{2u},\\
& C^{A_g}=\sigma_z, C^{B_{2g}}=i\sigma_y, C^{A_u}=I, C^{B_{2u}}=\sigma_x;\\
\end{split}
\end{equation}
and we choose $|0,0\rangle$ as the representation basis for $A_g$. Following the same procedure and dropping the constant, it leads to the Hamiltonian (\ref{txz hamiltonian}) with the expression:
\begin{equation}
H_{xz}=\sum_i H^1_{i}+H^{2,1}_{i,i+1}+H^{2,2}_{i}+H^3_{i,i+1}+H^4_{i,i+1}
\end{equation}

\begin{equation}
 H^1_{i} = \sum_{\tau} D(S^y_{\tau,i})^2,\nonumber\\
\end{equation}


\begin{equation}\nonumber
\begin{split}
& H^{2,1}_{i,i+1} = \sum_{\tau} W_1 \left[S^y_{\tau,i}S^y_{\tau,i+1} + (S^x_{\tau,i}S^x_{\tau,i+1} + S^z_{\tau,i}S^z_{\tau,i+1})^2\right]\\
&\quad\quad + W_2(S^y_{\tau,i}S^y_{\tau,i+1})^2 
 + W_3[(S^x_{\tau,i}S^x_{\tau,i+1})^2 + (S^z_{\tau,i}S^z_{\tau,i+1})^2]\\
&\quad\quad + W_4(S^x_{\tau,i}S^x_{\bar\tau,i+1} + S^z_{\tau,i}S^z_{\bar\tau,i+1})^2 
 + W_5[(S^x_{\tau,i}S^x_{\bar\tau,i+1})^2 + (S^z_{\tau,i}S^z_{\bar\tau,i+1})^2],\\
\end{split}
\end{equation}

\begin{equation}\nonumber
\begin{split}
& H^{2,2}_{i} = W_6(\mathbf{S}_{1,i}\cdot\mathbf{S}_{2,i})^2 
 + W_7[(S^x_{1,i}S^x_{2,i})^2 + (S^z_{1,i}S^z_{2,i})^2 - (S^y_{1,i}S^y_{2,i})^2],\\
\end{split}
\end{equation}

\begin{equation}\nonumber
\begin{split}
& H^3_{i,i+1} = U_1(\mathbf{S}_{1,i}\cdot\mathbf{S}_{2,i})^2\left[(S^y_{1,i+1})^2 + (S^y_{2,i+1})^2\right]
 + U_1\left[(S^y_{1,i})^2 + (S^y_{2,i})^2\right](\mathbf{S}_{1,i+1}\cdot\mathbf{S}_{2,i+1})^2\\
&\quad\quad + U_2(S^x_{1,i}S^z_{2,i})^2\left[(S^z_{1,i+1})^2 + (S^x_{2,i+1})^2\right]
 + U_2(S^z_{1,i}S^x_{2,i})^2\left[(S^x_{1,i+1})^2 + (S^z_{2,i+1})^2\right]\\
&\quad\quad + U_2\left[(S^z_{1,i})^2 + (S^x_{2,i})^2 \right](S^x_{1,i+1}S^z_{2,i+1})^2
 + U_2\left[(S^x_{1,i})^2 + (S^z_{2,i})^2 \right](S^z_{1,i+1}S^x_{2,i+1})^2\\
&\quad\quad + U_3[(S^x_{1,i}S^{xy}_{2,i} + S^z_{1,i}S^{yz}_{2,i} - S^{xy}_{1,i}S^x_{2,i} - S^{yz}_{1,i}S^z_{2,i})
(S^y_{1,i+1} - S^y_{2,i+1})]\\
&\quad\quad - U_3[(S^x_{1,i}S^z_{2,i} + S^z_{1,i}S^x_{2,i} + S^{xy}_{1,i}S^{yz}_{2,i} + S^{yz}_{1,i}S^{xy}_{2,i})
(S^{xz}_{1,i+1} + S^{xz}_{2,i+1})]\\
&\quad\quad + U_3[(S^y_{1,i} - S^y_{2,i})
 (S^x_{1,i+1}S^{xy}_{2,i+1} + S^z_{1,i+1}S^{yz}_{2,i+1} - S^{xy}_{1,i+1}S^x_{2,i+1} - S^{yz}_{1,i+1}S^z_{2,i+1})]\\
&\quad\quad - U_3[(S^{xz}_{1,i} + S^{xz}_{2,i})
 (S^x_{1,i+1}S^z_{2,i+1} + S^z_{1,i+1}S^x_{2,i+1} + S^{xy}_{1,i+1}S^{yz}_{2,i+1} + S^{yz}_{1,i+1}S^{xy}_{2,i+1})]\\
&\quad\quad + U_4\left[S^y_{1,i}(S^y_{2,i})^2 - (S^{y}_{1,i})^2S^y_{2,i}\right](S^y_{1,i+1} - S^y_{2,i+1})\\
&\quad\quad + U_4\left[S^{xz}_{1,i}(S^y_{2,i})^2 + (S^y_{1,i})^2S^{xz}_{2,i}\right](S^{xz}_{1,i+1} + S^{xz}_{2,i+1})\\
&\quad\quad + U_4(S^y_{1,i} - S^y_{2,i})\left[S^y_{1,i+1}(S^y_{2,i+1})^2 - (S^{y}_{1,i+1})^2S^y_{2,i+1}\right]\\
&\quad\quad + U_4(S^{xz}_{1,i} + S^{xz}_{2,i})\left[S^{xz}_{1,i+1}(S^y_{2,i+1})^2 + (S^y_{1,i+1})^2S^{xz}_{2,i+1}\right]\\
\end{split}
\end{equation}

\begin{equation}\nonumber
\begin{split}
& H^4_{i,i+1} = V_1(\mathbf{S}_{1,i}\cdot\mathbf{S}_{2,i})^2
[(S^x_{1,i+1}S^x_{2,i+1})^2 + (S^z_{1,i+1}S^z_{2,i+1})^2 - (S^y_{1,i+1}S^y_{2,i+1})^2]\\
& \quad\quad + V_1[(S^x_{1,i}S^x_{2,i})^2 + (S^z_{1,i}S^z_{2,i})^2 - (S^y_{1,i}S^y_{2,i})^2]
 (\mathbf{S}_{1,i+1}\cdot\mathbf{S}_{2,i+1})^2\\
& \quad\quad + V_2\left[(S^z_{1,i}S^x_{2,i}S^x_{1,i+1}S^z_{2,i+1})^2
 + (S^x_{1,i}S^z_{2,i}S^z_{1,i+1}S^x_{2,i+1})^2\right]\\
&\quad\quad + V_3(\mathbf{S}_{1,i}\cdot\mathbf{S}_{2,i})^2(\mathbf{S}_{1,i+1}\cdot\mathbf{S}_{2,i+1})^2\\
&\quad\quad + V_4(S^y_{1,i}S^y_{2,i} + S^{xz}_{1,i}S^{xz}_{2,i})(S^y_{1,i+1}S^y_{2,i+1} + S^{xz}_{1,i+1}S^{xz}_{2,i+1})\\
&\quad\quad + V_4(S^y_{1,i}S^{xz}_{2,i} - S^{xz}_{1,i}S^y_{2,i})(S^y_{1,i+1}S^{xz}_{2,i+1} - S^{xz}_{1,i+1}S^y_{2,i+1})\\
&\quad\quad + \sum_{\tau}[V_5(S^z_{\tau,i}S^x_{\bar\tau,i} + S^{xy}_{\tau,i}S^{yz}_{\bar\tau,i}) + V_6(S^x_{\tau,i}S^z_{\bar\tau,i} + S^{yz}_{\tau,i}S^{xy}_{\bar\tau,i})\\
&\quad\quad + V_7(S^{xz}_{\tau,i}(S^x_{\bar\tau,i})^2 + (S^z_{\tau,i})^2S^{xz}_{\bar\tau,i})
 + V_8(S^{xz}_{\tau,i}(S^z_{\bar\tau,i})^2 + (S^{x}_{\tau,i})^2S^{xz}_{\bar\tau,i})]\cdot\\
& \quad\quad (S^x_{\tau,i+1}S^z_{\bar\tau,i+1} + S^{yz}_{\tau,i+1}S^{xy}_{\bar\tau,i+1})\\
& \quad\quad + [V_5(S^z_{\tau,i}S^{yz}_{\bar\tau,i} - S^{xy}_{\tau,i}S^x_{\bar\tau,i}) +
V_6(S^x_{\tau,i}S^{xy}_{\bar\tau,i} - S^{yz}_{\tau,i}S^{z}_{\bar\tau,i})\\
&\quad\quad + V_7((S^z_{\tau,i})^2S^y_{\bar\tau,i} - S^y_{\tau,i}(S^x_{\bar\tau,i}))^2
 + V_8((S^x_{\tau,i})^2S^y_{\bar\tau,i} - S^y_{\tau,i}(S^z_{\bar\tau,i})^2)]\cdot\\
& \quad\quad (S^x_{\tau,i+1}S^{xy}_{\bar\tau,i+1} - S^{yz}_{\tau,i+1}S^z_{\bar\tau,i+1})\\
& \quad\quad + [V_9(S^{xy}_{\tau,i}S^x_{\bar\tau,i} - S^z_{\tau,i}S^{yz}_{\bar\tau,i}) +
 V_{10}(S^{yz}_{\tau,i}S^z_{\bar\tau,i} - S^x_{\tau,i}S^{xy}_{\bar\tau,i})\\
& \quad\quad + V_{11}(S^y_{\tau,i}(S^z_{\bar\tau,i})^2 - (S^x_{\tau,i})^2S^y_{\bar\tau,i})
 + V_{12}(S^y_{\tau,i}(S^x_{\bar\tau,i})^2 - (S^z_{\tau,i})^2S^y_{\bar\tau,i})]\cdot\\
&\quad\quad (S^y_{\tau,i+1}(S^x_{\bar\tau,i+1})^2 - (S^z_{\tau,i+1})^2S^y_{\bar\tau,i+1})\\
&\quad\quad + [V_9(S^x_{\tau,i}S^z_{\bar\tau,i} + S^{yz}_{\tau,i}S^{xy}_{\bar\tau,i}) +
V_{10}(S^z_{\tau,i}S^x_{\bar\tau,i} + S^{xy}_{\tau,i}S^{yz}_{\bar\tau,i})\\
&\quad\quad + V_{11}(S^{xz}_{\tau,i}(S^x_{\bar\tau,i})^2 + (S^z_{\tau,i})^2S^{xz}_{\bar\tau,i})
 + V_{12}(S^{xz}_{\tau,i}(S^z_{\bar\tau,i})^2 + (S^x_{\tau,i})^2S^{xz}_{\bar\tau,i})]\cdot\\
&\quad\quad((S^x_{\tau,i+1})^2S^{xz}_{\bar\tau,i+1} + S^{xz}_{\tau,i+1}(S^z_{\bar\tau,i+1})^2)\\
\end{split}
\end{equation}

where the parameters are given as: $D = \frac{14}{3},W_1=W_4= -\frac{5}{9},W_2=-\frac{5}{6},W_3=\frac{25}{18},W_5=-\frac{5}{18},\ W_6=-\frac{16}{27}, W_7=\frac{4}{3}, U_1=\frac{1}{6}, U_2=\frac{5}{6}, U_3=\frac{5}{72},U_4=\frac{5}{36},\ V_1=\frac{1}{6}, V_2=-\frac{5}{6}, V_3=-\frac{1}{27}, V_4=V_6=\frac{1}{48}, V_5=\frac{1}{72},\ V_7=V_{10}=\frac{1}{24},V_8=V_9=\frac{1}{36}, V_{11}=\frac{1}{18}, V_{12}=\frac{1}{12},$
and $\tau=1,2,\bar\tau=3-\tau$ label the two chains, $S^{mn}=S^mS^n+S^nS^m(m,n=x,y,z, \mathrm{with} \quad m\neq n)$.

The active operators have been discussed in Ref. \cite{LiuLiuWen} and will not be repeated here.


\begin{thebibliography}{99}
\bibitem{note2}
Note that, for the line $\theta=\frac{\pi}{2}$, the Hamiltonian (\ref{2d_PD_Ham}) has an enlarged $U(1)$ symmetry generated by $S_y^{+}$. Thus the magnetic order along this line is owing to finite $D$ and should vanish in the $D\to\infty$ limit.
\bibitem{note1}
It should be mentioned that near $\lambda=0.9$, there is a very small region showing nonzero expectation values of $|\langle S_{i}^{y+}\rangle|$ and $|\langle S_{i}^{z+}\rangle|$, while $|\langle S_{i}^{y-}\rangle|$ and $|\langle S_{i}^{z-}\rangle|$ are non zero near $\lambda=0.1$ (not shown in the figure). Their order of magnitude is $10^{-2}$. These orders break the $U(1)$ symmetry and is owing to finite $D$, which will vanish in the limit $D\to\infty$.
\end{thebibliography}
\end{document}